\documentclass[aip, apl, reprint, superscriptaddress, amsmath,amssymb, floatfix]{revtex4-1}

\usepackage{graphicx}
\usepackage{amsmath}

\begin{document}

%\preprint{}

\title{Interfacing topological insulators and ferrimagnets: Bi$_2$Te$_3$ and Fe$_3$O$_4$ heterostructures grown by molecular beam epitaxy}


\author{V.M. Pereira}
\email{vanda.pereira@cpfs.mpg.de}
\affiliation{Max Planck Institute for Chemical Physics of Solids, N{\"o}thnitzer Str. 40, 01187 Dresden, Germany}

\author{C.N. Wu}
\affiliation{Max Planck Institute for Chemical Physics of Solids, N{\"o}thnitzer Str. 40, 01187 Dresden, Germany}
\affiliation{Department of Physics, National Tsing Hua University, Hsinchu 30013, Taiwan}

\author{C.-A. Knight}
\affiliation{Max Planck Institute for Chemical Physics of Solids, N{\"o}thnitzer Str. 40, 01187 Dresden, Germany}
\affiliation{Department of Physics and Astronomy, University of British Columbia, Vancouver, BC V6T 1Z1, Canada}

\author{A. Choa}
\affiliation{Max Planck Institute for Chemical Physics of Solids, N{\"o}thnitzer Str. 40, 01187 Dresden, Germany}
\affiliation{Department of Physics and Astronomy, University of British Columbia, Vancouver, BC V6T 1Z1, Canada}

\author{L.H. Tjeng}
\affiliation{Max Planck Institute for Chemical Physics of Solids, N{\"o}thnitzer Str. 40, 01187 Dresden, Germany}

\author{S.G. Altendorf}
\affiliation{Max Planck Institute for Chemical Physics of Solids, N{\"o}thnitzer Str. 40, 01187 Dresden, Germany}


\date{\today}

\begin{abstract}
Relying on the magnetism induced by the proximity effect in heterostructures of topological insulators and magnetic insulators is one of the promising routes to achieve the quantum anomalous Hall effect. Here we investigate heterostructures of Bi$_2$Te$_3$ and Fe$_3$O$_4$. By growing two different types of heterostructures by molecular beam epitaxy, Fe$_3$O$_4$ on Bi$_2$Te$_3$ and Bi$_2$Te$_3$ on Fe$_3$O$_4$, we explore differences in chemical stability, crystalline quality, electronic structure, and transport properties. We find the heterostructure Bi$_2$Te$_3$ on Fe$_3$O$_4$ to be a more viable approach, with transport signatures in agreement with a gap opening in the topological surface states.

\end{abstract}

%\keywords{}

\maketitle


\section{Introduction}

Since its initial theoretical prediction \cite{Fu2007}, topological insulators (TIs) like the prototypical Bi$_2$Te$_3$ and Bi$_2$Se$_3$ have been extensively studied due to the multitude of features stemming from their topological surface states. Many of the recent studies on TIs have concentrated on breaking the time reversal symmetry by introducing magnetic order in the system, as this can lead to exotic phenomena such as the quantum anomalous Hall effect (QAHE). Requisite to this is the opening of a gap in the topological surface states, which can be experimentally achieved either by magnetic doping of the TI \cite{Chang2013, Chang2015, Mogi2015, Ye2015} or by making use of the magnetic proximity effect at the interface between a TI and a magnetic layer \cite{Wei2013, Jiang2014, Tang2017, Yang2019}. Although magnetic doping has been proven to be an effective approach and the QAHE has been observed for these systems \cite{Chang2013}, the magnetic proximity effect can carry significant advantages. In addition to the much higher Curie temperature of the magnetic layer, one can avoid the inherent inhomogeneity of the doping process and additional scattering processes that reduce the mobility. One can also achieve a uniform magnetization at the interface between the TI and the magnetic layer. This could lead to an increase of the temperature for which the QAHE is observed, surpassing the low temperatures of less than 2 K \cite{Chang2013, Chang2015, Mogi2015} that are currently necessary for magnetically doped TI systems.

Earlier studies of interfaces between TIs and ferromagnets focused on the use of Fe as an adlayer \cite{Wray2011, He2011, Honolka2012, Polyakov2015, Sanchez-Barriga2017}. However, several studies found that the interface of Fe/TI is not clean, and that the  Fe atoms penetrate into the TI layer, forming an interface layer containing FeTe or FeSe \cite{Polyakov2015, Sanchez-Barriga2017}. Since any possible applications of these heterostructures for spintronic devices rely on well-defined interfaces, it is crucial to have a sharp interface, without chemical reactions between the layers. For this purpose, magnetic transition metal oxide insulators are promising candidates, due to their relatively inert nature when compared to the magnetic transition metals themselves. Being insulators, these will not contribute to the conductivity and, as a result, unambiguous monitoring of the topological surface states can be readily achieved.

In this work, we have selected Fe$_3$O$_4$ (magnetite) and Bi$_2$Te$_3$ as our respective magnetic layer and TI. Magnetite is an extensively investigated ferrimagnet owing to its interesting electrical and magnetic properties, from which we emphasize the high Curie temperature around 860 K. It also has a characteristic first-order metal-insulator transition occurring at 124 K, known as the Verwey transition \cite{Verwey1939}. We have chosen Fe$_3$O$_4$ because the detailed growth conditions, needed to obtain high quality films, have been recently established by Liu \textit{et al.} \cite{Liu2014, Liu2016}. Bi$_2$Te$_3$ was selected as the TI because it has been demonstrated by H{\"o}fer \textit{et al.} \cite{Hoefer2014} that stoichiometric films of Bi$_2$Te$_3$ can be grown, which are truly insulating in their bulk and show only intrinsic surface conductivity. Here we investigate the interface between these materials. We study the transport properties of the heterostructures to detect the presence of the proximity effect. The opening of an exchange gap in the TI surface states will lead to a suppression of the weak anti-localization (WAL) effect, a characteristic of TIs, and will induce the weak localization (WL) effect \cite{Lu2011, Kandala2013, Jiang2014, Yang2019}. Here we note that for the case of Bi$_2$Te$_3$ the QAHE can be induced not only by an out-of-plane magnetization but also by an in-plane magnetization due to the warping effects of the topological surface states \cite{Fu2009, Liu2013}, giving us more flexibility in the choice of the magnetic material and interface.

The present work reports on two different heterostructures of Bi$_2$Te$_3$ and Fe$_3$O$_4$, namely Fe$_3$O$_4$ / Bi$_2$Te$_3$ / Al$_2$O$_3$ (0001) and Bi$_2$Te$_3$ / Fe$_3$O$_4$ / MgO (001). These were characterized \textit{in situ} by reflection high-energy electron diffraction, x-ray photoelectron spectroscopy, and angle-resolved photoelectron spectroscopy. We then present the transport properties of these heterostructures and discuss the findings.

\section{Experiment}

\normalsize
The films were prepared by molecular beam epitaxy (MBE) in an \textit{in situ} system with base pressure of about 2$\times$10$^{-10}$ mbar. This comprises two MBE growth chambers, one dedicated to Bi$_2$Te$_3$ and the other to Fe$_3$O$_4$.
 
For the investigation of Fe$_3$O$_4$ on Bi$_2$Te$_3$, the heterostructure described in the first part of this manuscript, Bi$_2$Te$_3$ films were grown on Al$_2$O$_3$ (0001) substrates purchased from Crystec GmbH. Prior to the deposition, the substrates were annealed at 600 $^\circ$C for 2 hours in an oxygen pressure of 1$\times$10$^{-6}$ mbar. High purity Bi and Te were evaporated from  effusion cells with flux rates measured by a quartz crystal monitor at the growth position. The flux rates were set at 0.5 \AA/min for Bi ($T_{\rm{Bi}} \approx$ 458 $^\circ$C) and 1.5 \AA/min for Te ($T_{\rm{Te}} \approx$ 228 $^\circ$C). Bi$_2$Te$_3$ was grown in a two-step procedure. First, 3 quintuple layers (QLs) were deposited at 160 $^\circ$C and annealed at 240 $^\circ$C in Te atmosphere for 30 minutes. Then, we deposited 7 more QLs at 220 $^\circ$C, amounting to a final thickness of 10 QLs (see also Ref. \onlinecite{Hoefer2014}). Subsequently, Fe was evaporated at temperatures of about 1185 - 1230 $^\circ$C (flux rate of Fe, $\phi_{\rm{Fe}}$ = 0.5 - 1 \AA/min) in a pure oxygen atmosphere on top of the Bi$_2$Te$_3$ layer. Molecular oxygen was supplied through a leak valve, varying the partial pressure, P$_{\rm{Ox}}$, between 5$\times$10$^{-8}$ and 1$\times$10$^{-5}$ mbar.  

For the heterostructure with Bi$_2$Te$_3$ on Fe$_3$O$_4$, described in the second part, 30 nm-thick Fe$_3$O$_4$ films were grown on MgO (001) substrates purchased from Crystec GmbH, following the recipe from Refs. \onlinecite{Liu2014, Liu2016, Chang2016}. Subsequently, Bi$_2$Te$_3$ was deposited on top, using a slightly modified two-step procedure. 

Reflection high-energy electron diffraction (RHEED) was used to monitor in \textit{real-time} the epitaxial growth, using a STAIB Instruments RH35 system with the kinetic energy of the electrons set at 15 keV (Bi$_2$Te$_3$) or 20 keV (Fe$_3$O$_4$). 
All samples were characterized \textit{in situ} by x-ray photoelectron spectroscopy (XPS) using monochromatized Al K$_{\alpha}$ light (1486.6 eV) and angle-resolved photoelectron spectroscopy (ARPES) using a non-monochromatic He discharge lamp with 21.2 eV photon energy (He I line) at room temperature and using a Scienta R3000 electron energy analyzer.

In order to characterize the structural quality of the films, \textit{ex situ} x-ray diffraction (XRD) measurements were performed with a PANalytical X’Pert PRO diffractometer using monochromatic Cu-K$_{\alpha 1}$ radiation ($\lambda$ = 1.54056 \AA). Atomic force microscopy (AFM) was carried out using a Veeco Metrology MultiMode Atomic Force Microscope (Model 920-006-101) in tapping mode.

To avoid the contamination of Bi$_2$Te$_3$ during \textit{ex situ} transport measurements, the heterostructure with Bi$_2$Te$_3$ on Fe$_3$O$_4$ was capped \textit{in situ} with 12 nm of Te grown at room temperature \cite{Hoefer2015} prior to the transfer to outside of the UHV system. Electrodes and connections were made of cut and pressed indium balls and copper wires in a standard Van der Pauw configuration. Transport measurements were performed using a Physical Property Measurement System from Quantum Design with a base temperature of 2 K.

\begin{figure}[htbp]
\includegraphics[width=0.75\columnwidth]{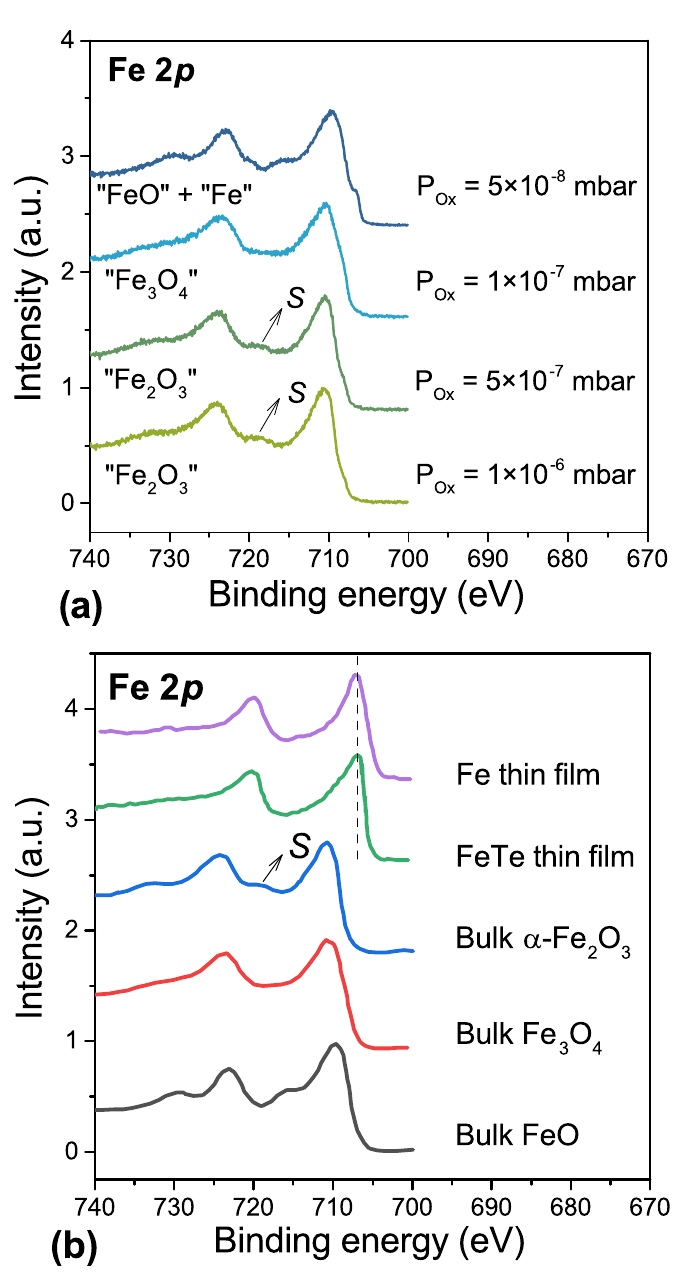}
\caption{(a) Fe 2\textit{p} XPS spectra of nominally 40-nm-thick FeO$_x$ films grown on 10 QL Bi$_2$Te$_3$ at 50-55 $^\circ$C substrate temperature and $\phi_{\rm{Fe}}$ = 1 \AA/min under various oxygen pressures. The resulting phase is indicated by the labels in quotation marks. (b) Reference Fe 2\textit{p} XPS spectra of Fe and FeTe thin films (reproduced from Ref. \onlinecite{Telesca2012}) and bulk $\alpha$-Fe$_2$O$_3$, Fe$_3$O$_4$ and FeO (reproduced from Ref. \onlinecite{Gota1999}).} 
\label{Fig1}
\end{figure}

\begin{figure*}[hbtp]
\includegraphics[width=\textwidth]{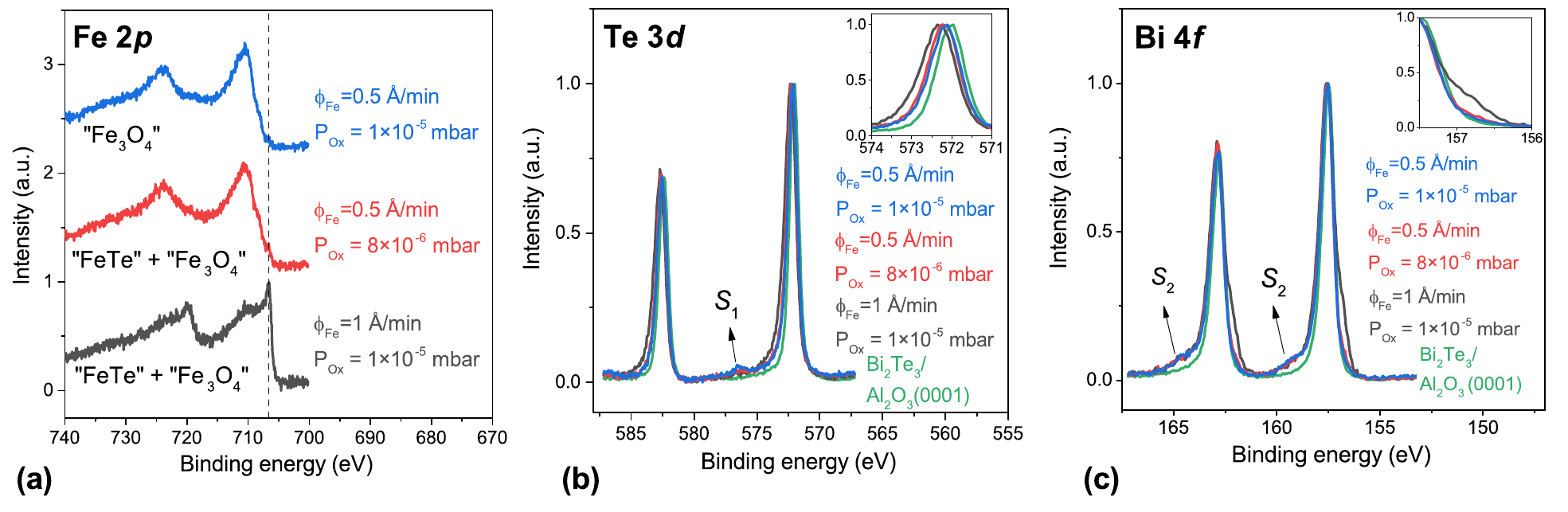}
\caption{XPS spectra of the Fe 2\textit{p} (a), Te 3\textit{d} (b) and Bi 4\textit{f} (c) core levels for various ratios of Fe to O$_2$, for a nominal thickness of 6 MLs of FeO$_x$ on top of 10 QLs of Bi$_2$Te$_3$. The resulting phases are indicated by the labels in quotation marks. For (b) and (c), the reference XPS spectra of a 10 QL Bi$_2$Te$_3$ thin film grown on Al$_2$O$_3$ (0001) are shown in green. The insets show a closeup of the Te 3\textit{d} and Bi 4\textit{f} peaks.} 
\label{Fig2}
\end{figure*}

\begin{figure*}[htbp]
\includegraphics[width=\textwidth]{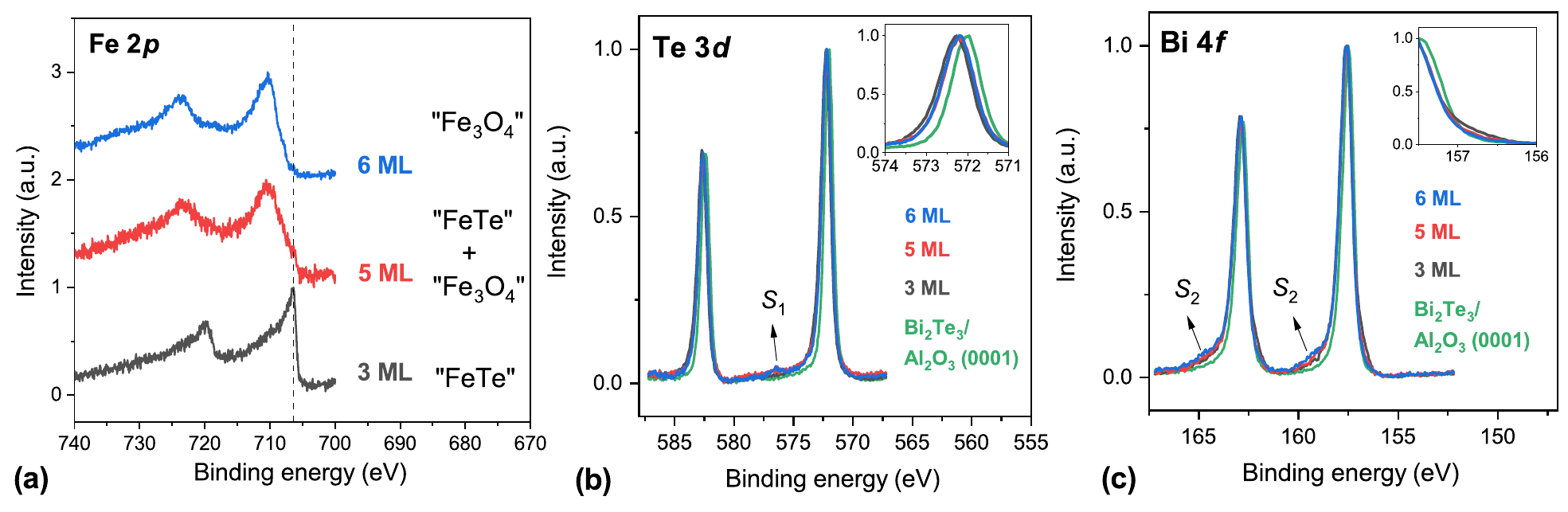}
\caption{XPS spectra of the Fe 2\textit{p} (a), Te 3\textit{d} (b) and Bi 4\textit{f} (c) core level for various nominal thicknesses of FeO$_x$ on top of 10 QLs of Bi$_2$Te$_3$. The Fe flux rate and oxygen pressure were kept constant at 0.5 \AA/min and 1$\times 10^{-5}$ mbar, respectively. The resulting phases are indicated by the labels in quotation marks. For (b) and (c), the reference XPS spectra of a 10 QL Bi$_2$Te$_3$ thin film grown on Al$_2$O$_3$ (0001) are shown in green.} 
\label{Fig3}
\end{figure*}

\section{Results and discussion} 
\subsection{Fe$_3$O$_4$ on Bi$_2$Te$_3$ on Al$_2$O$_3$ (0001)}

For all the films in the current subsection, 10 QLs of Bi$_2$Te$_3$ were deposited onto Al$_2$O$_3$ (0001) substrates as described in the Experiment section. Afterwards, Fe was deposited in an oxygen atmosphere, varying the growth parameters such as  the flux rate of Fe and the oxygen partial pressure. The parameters were optimized in order to avoid strong chemical reactions between the layers and ultimately achieve a satisfactory growth of Fe$_3$O$_4$. For all the films reported here, the substrate temperature ranged from 50-55 $^\circ$C due to the radiation heat load from the Fe effusion cell in thermal equilibrium with the unheated sample holder. All attempts to grow Fe$_3$O$_4$ at higher temperatures led to strong chemical reactions with the TI layer.

Our first aim was to establish the optimal conditions required for a sustained growth of Fe$_3$O$_4$ films at this substrate temperature range. We began with the conditions that were found to be optimal for films grown on oxide substrates at 250 $^\circ$C (Refs. \onlinecite{Liu2014, Liu2016, Chang2016}), namely an Fe flux rate of 1 \AA/min with an oxygen partial pressure of 1$\times$10$^{-6}$ mbar. The bottom (light green) curve of Figure \ref{Fig1}(a) shows the Fe 2$p$ XPS spectrum of a nominally 40 nm FeO$_x$ film grown under these conditions on 10 QLs of Bi$_2$Te$_3$. In Figure \ref{Fig1}(b) we have also collected Fe 2$p$ spectra of several reference Fe compounds, which include an Fe metal film (purple curve), FeTe film (green), Fe$_2$O$_3$ bulk (blue), Fe$_3$O$_4$ bulk (red), and FeO bulk (dark gray). We observe from the spectral line shape that the 40 nm FeO$_x$ film (Fig. \ref{Fig1}(a) light green) has all the characteristics of a Fe$_2$O$_3$-like phase (Fig. \ref{Fig1}(b) blue), i.e. it is overoxidized, as is indicated by the presence of the satellite peak labeled $S$. We then lowered the oxygen pressure to 5$\times$10$^{-7}$ mbar and obtained a 40 nm FeO$_x$ film (Fig. \ref{Fig1}(a) dark green) that is also Fe$_2$O$_3$-like, with the $S$ feature still present. Lowering the pressure even further to 1$\times$10$^{-7}$ mbar produces a film for which the Fe 2$p$ spectrum (Fig. \ref{Fig1}(a) light blue) is quite similar to that of Fe$_3$O$_4$ bulk (Fig. \ref{Fig1}(b) red). Finally, with 5$\times$10$^{-8}$ mbar pressure, the Fe 2$p$ spectrum of the film (Fig. \ref{Fig1}(a) dark blue) shows features that belong to FeO (Fig. \ref{Fig1}(b) dark gray) and Fe metal (Fig. \ref{Fig1}(b) purple). These results thus present a significant difference from the ones reported for the growth of magnetite on oxide substrates at temperatures of 250 $^\circ$C (Refs. \onlinecite{Liu2014, Liu2016, Chang2016}). It appears that high substrate temperatures allow for a wide range of oxygen pressures leading to the formation of good quality magnetite films, while the lower temperature necessary for the growth on Bi$_2$Te$_3$ considerably narrows the growth window of magnetite.

In the next step, we investigated the growth process of FeO$_x$ closer to the interface with the Bi$_2$Te$_3$ layer. To this end we prepared a series of samples with nominally 6 monolayers (MLs) of FeO$_x$ using various ratios of Fe to oxygen (here given by the flux rate of Fe and oxygen partial pressure). For these thinner films, we observed that the relatively low oxygen pressures, similar to the ones used previously, led to strong Fe-Te reactions at the interface indicating the need for higher oxygen pressures for the growth of the Fe$_3$O$_4$ layer. However, even for a pressure of 1$\times$10$^{-5}$ mbar, we also observe that the Fe 2$p$ spectrum of the resulting 6 ML film, as shown by the bottom curve (dark gray) in Figure \ref{Fig2}(a), contains features that belong to a mixture of Fe$_3$O$_4$ and Fe metal and/or FeTe. Here we emphasize the presence of the intense peak at 707 eV binding energy, marked by the dashed line in Figure \ref{Fig2}(a), that is characteristic for Fe metal or FeTe, indicated also by the dashed line in Fig. \ref{Fig1}(b). A further comparison of the corresponding Te 3\textit{d} and Bi 4\textit{f} spectra with the reference 10 QL Bi$_2$Te$_3$ grown on Al$_2$O$_3$ (green curves) in Figures \ref{Fig2}(b) and (c) shows a shift towards higher binding energies and a shoulder at lower binding energies, respectively, as seen in the insets. These results strongly indicate the presence of a reaction between both layers, leading to the formation of FeTe and the consequent appearance of metallic bismuth at 157 and 162 eV (Ref. \onlinecite{Moulder1992}).  

Subsequently, we further increased the oxygen to Fe ratio. The red curve in Figure \ref{Fig2} uses an Fe flux rate of 0.5 \AA/min and oxygen partial pressure of 8$\times$10$^{-6}$ mbar. The Fe 2\textit{p} spectrum shows a considerable reduction of the peak at 707 eV, indicating less reaction between the FeO$_x$ and Bi$_2$Te$_3$ layers. Indeed, the Te 3\textit{d} and Bi 4\textit{f} spectra are now more similar to the reference. Finally, the film prepared with Fe flux rate of 0.5 \AA/min and oxygen partial pressure of 1$\times$10$^{-5}$ mbar (blue curve) presents the Fe 2\textit{p} spectrum with much more similar features to the one expected for Fe$_3$O$_4$ (see Figure \ref{Fig1}(b), red curve). The peak at 707 eV is no longer visible and no extra satellite peaks (indicative of the formation of FeO and Fe$_2$O$_3$ phases, for instance) can be observed. In addition to this, the Te 3\textit{d} and Bi 4\textit{f} core levels are more similar to the expected for Bi$_2$Te$_3$, indicating that the reactions between the layers are minimized. However, it should be noted that small amounts of tellurium and bismuth oxides are formed, as indicated by the small peaks/shoulders (\textit{S$_1$} and \textit{S$_2$}) observed at higher binding energies on the Te 3\textit{d} and Bi 4\textit{f} spectra.

To obtain more information about the intricacies of the interface between FeO$_x$ and Bi$_2$Te$_3$, we also studied the effect of the thickness of the Fe oxide layer for the thinnest films. To this end, the flux rate and oxygen partial pressure were kept constant, following the best results found in the previous experiment ($\phi_{\rm{Fe}}$ = 0.5 \AA/min and P$_{\rm{Ox}}$ = 1$\times$10$^{-5}$ mbar) and the nominal thickness was varied between 3 and 6 MLs. The Fe 2\textit{p} XPS spectra are plotted in Figure \ref{Fig3}(a). Rather than just the Fe/O$_2$ ratio, it can be noticed that also the nominal thickness is relevant for the successful growth of Fe$_3$O$_4$ on top of Bi$_2$Te$_3$. For very thin layers (gray curve), we observe a predominance of the Fe-Te bond (dashed line), while the Te 3\textit{d} and Bi 4\textit{f} spectra show similar trends to those reported in Figure \ref{Fig2}, indicating the formation of FeTe and metallic Bi at the interface. It is possible that a significantly higher ratio of Fe to oxygen is needed in order to obtain Fe$_3$O$_4$ for such thin layers. However, the applied pressure is already near the limit of what can be tolerated in the MBE system. Nevertheless, as the nominal thickness is increased, Fe oxides start to form and, for 5 MLs, the presence of the peak at 707 eV is substantially reduced. Lastly, for 6 MLs the measured spectrum is very similar to that of Fe$_3$O$_4$. We note that the probing depth for 1486.6 eV photons is larger than the thicknesses of the films used here. Therefore, the signal from the first 3 MLs (Fe-Te bond) should still be observable for films with 5 and 6 MLs. The absence of the FeTe signal for thicker layers suggests that the thickness – for a constant Fe/O$_2$ ratio – is indeed one of the controlling factors.

From the results of this study, we conclude that the growth of Fe$_3$O$_4$ on top of Bi$_2$Te$_3$ is possible, but not perfect. To keep the reactions at the interface minimized, a two-step procedure was implemented: growing first 6 MLs with $\phi_{\rm{Fe}}$ = 0.5 \AA/min and P$_{\rm{Ox}}$ = 1$\times$10$^{-5}$ mbar and then growing up to 40 nm using an Fe flux rate of $\phi_{\rm{Fe}}$ = 1 \AA/min and lower oxygen pressure, P$_{\rm{Ox}}$ = 1$\times$10$^{-7}$ mbar, at a substrate temperature of roughly 50-55 $^\circ$C. 

\begin{figure}[htbp]
\includegraphics[width=\columnwidth]{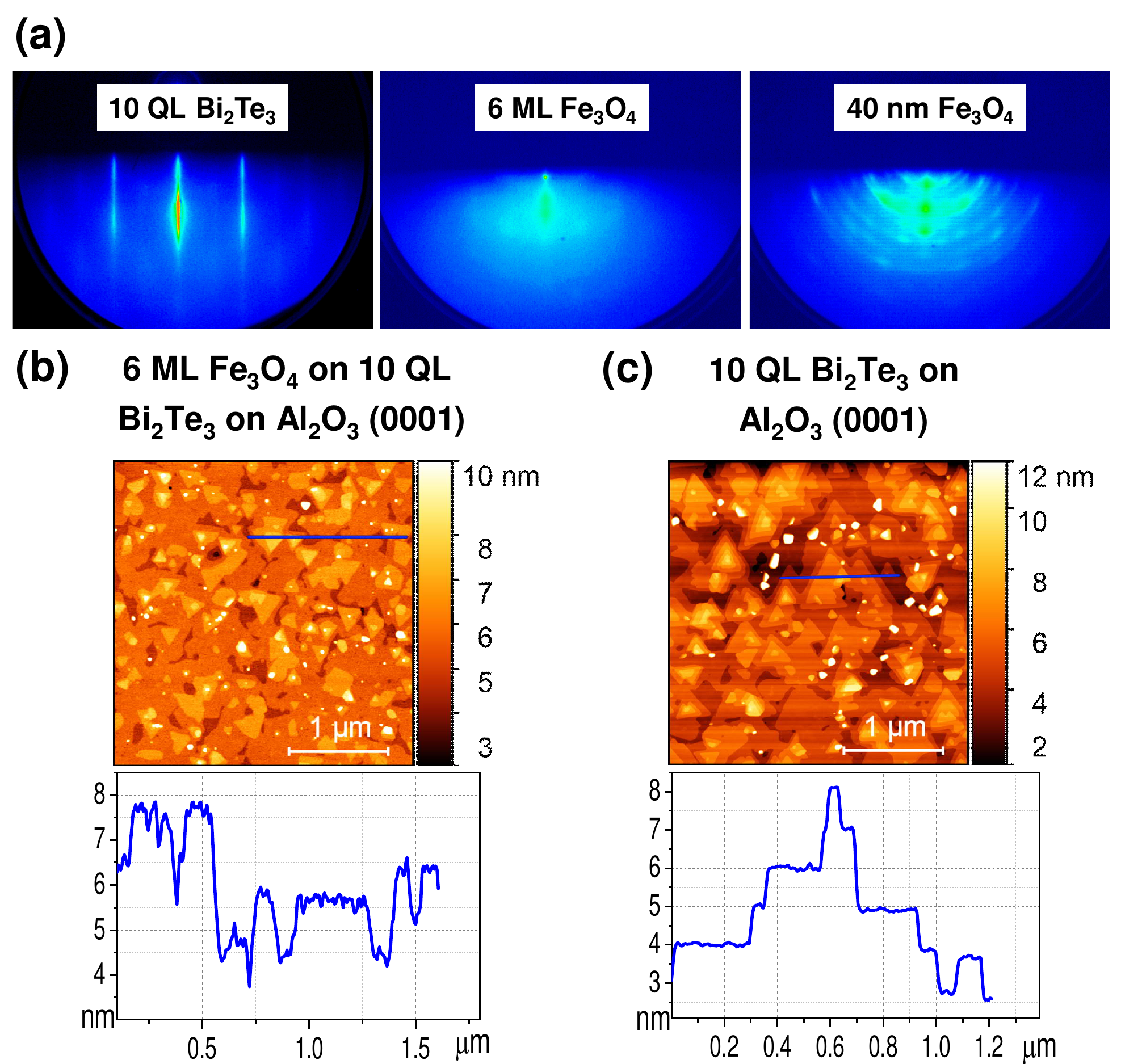}
\caption{(a) RHEED patterns of each step of the heterostructure: 10 QLs of Bi$_2$Te$_3$ on Al$_2$O$_3$ (0001) \textit{(left)}; 6 MLs of Fe$_3$O$_4$ grown with $\phi_{\rm{Fe}}$ = 0.5 \AA/min and P$_{\rm{Ox}}$ = 1$\times$10$^{-5}$ mbar \textit{(center)} and 40 nm of Fe$_3$O$_4$ with $\phi_{\rm{Fe}}$ = 1 \AA/min and P$_{\rm{Ox}}$ = 1$\times$10$^{-7}$ mbar \textit{(right)}. (b) Morphological characterization by AFM of 6 MLs of Fe$_3$O$_4$ on 10 QLs of Bi$_2$Te$_3$ on Al$_2$O$_3$ (0001), (c) 10 QLs of Bi$_2$Te$_3$ on Al$_2$O$_3$ (0001). The blue lines in the AFM pictures represent the locations of the height profiles plotted below.} \label{Fig4}
\end{figure}

Figure \ref{Fig4}(a) depicts the RHEED patterns for the Bi$_2$Te$_3$ layer and the Fe oxide overlayer after each of the two steps of the growth. The streaky lines noticeable for 10 QLs of Bi$_2$Te$_3$ show the good quality of the topological insulator layer. Upon the growth of 6 MLs of magnetite, we have a predominance of an amorphous background where some lines/spots can be distinguished. For 40 nm of Fe$_3$O$_4$, the pattern shows rings and spots indicating a polycrystalline and possibly 3D growth. 

\begin{figure*}[htbp]
\includegraphics[width=\textwidth]{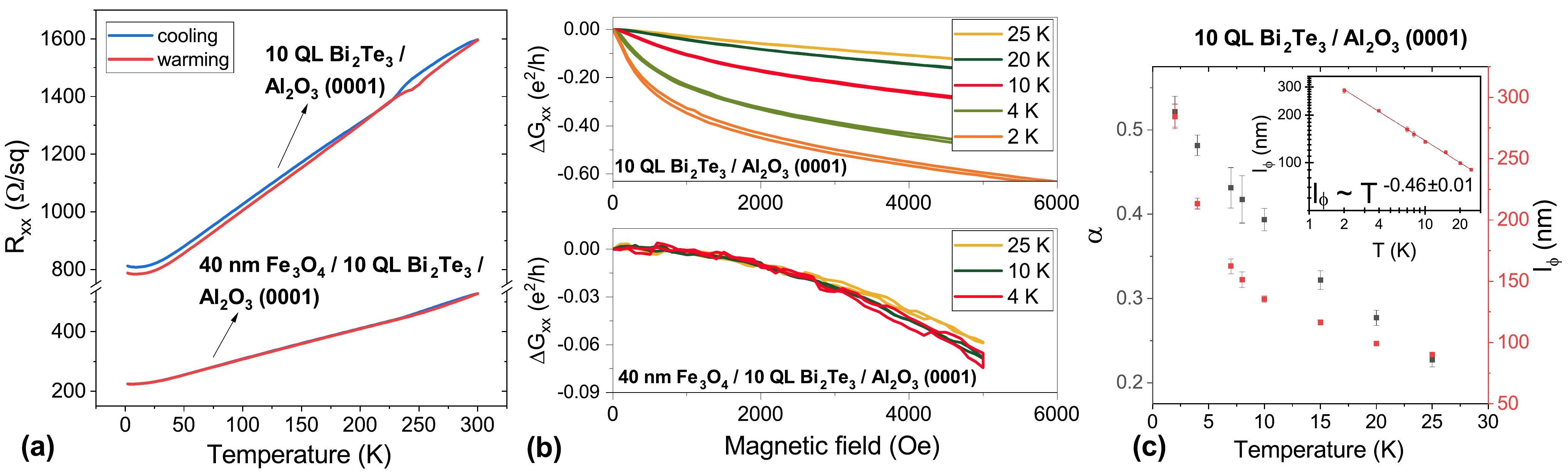}
\caption{(a) Sheet resistance as a function of temperature for 10 QLs of Bi$_2$Te$_3$ on Al$_2$O$_3$ (0001) and for the optimized heterostructure of 40 nm of Fe$_3$O$_4$ grown on top of 10 QLs of Bi$_2$Te$_3$. (b) Magnetoconductance for both samples. For the heterostructure, the magnetoconductance shows only a parabolic $B$-field dependence. (c) Dependence of the HLN fitting parameters, $\alpha$ and $l_{\phi}$, for the TI grown on Al$_2$O$_3$ (0001).} \label{Fig5}
\end{figure*}

The morphological characterization of 6 MLs of Fe$_3$O$_4$ on 10 QLs of Bi$_2$Te$_3$ on Al$_2$O$_3$ (0001) is depicted in Figure \ref{Fig4}(b), as well as a reference TI film with 10 QLs (c). The AFM pictures show that the magnetite layer covers the Bi$_2$Te$_3$ in a relatively uniform manner, and the pyramid structures with 1 QL-steps, typical of Bi$_2$Te$_3$, can still be observed.

To search for the possible presence of the magnetic proximity effect, we conducted temperature-dependent resistance measurements, depicted in Figure \ref{Fig5}(a). The same figure also shows the sheet resistance of a 10 QL TI thin film grown on Al$_2$O$_3$ (0001) for comparison. Both curves present quite similar features, with a metallic-like behavior, typical of the topological surface states, being predominant over the majority of the temperature range; for low temperatures, an upturn characteristic of TIs is observed. However, the absolute value of the sheet resistance diminishes considerably for the heterostructure. The formation of bismuth and tellurium oxides, observed by XPS, and some residual FeTe at the interface, can lead to doping of the TI due to Te vacancies and anti-site defects, thus increasing the contribution of the bulk to the transport properties and decreasing the overall resistance. Moreover, considering a parallel resistance between the layers of Bi$_2$Te$_3$ and Fe$_3$O$_4$, it would be expected that the sheet resistance would be dominated by the magnetite signal above the Verwey transition temperature (for 40 nm Fe$_3$O$_4$ on MgO (001), the sheet resistance is $\approx$ 1200 $\Omega$/sq at room temperature and the Verwey transition occurs at $T_{\rm{V}}\approx$ 119 K, cf. Ref. \onlinecite{Liu2014}). When $T < T_{\rm{V}}$, the magnetite layer is expected to be much more insulating and therefore the TI should be the main contributor to the resistance. The absence of the Verwey transition in this heterostructure, combined with the low crystalline order observed from RHEED, hints towards a subpar quality of the magnetite layer. 

One possibility is that Bi and/or Te constituents are incorporated into the magnetite, leading to the formation of doped Fe$_3$O$_4$ and therefore suppressing the characteristic Verwey transition. Furthermore, the XPS sensitivity is a limiting factor on the identification of phases. It is also conceivable that the magnetite overlayer might contain small amounts of parasitic phases correspondent to other iron oxides, as FeO and Fe$_2$O$_3$, which are below the detection limit of this technique. On the other hand, the absence of a clear Verwey transition has been frequently reported in the literature for Fe$_3$O$_4$ films \cite{Liu2014, Eerenstein2002}. Liu \textit{et al.} have investigated this phenomenon and found out that the Verwey transition temperature is strongly dependent on the size of the Fe$_3$O$_4$ crystallites, i.e. it starts to rapidly decrease if the crystallite or domain size becomes smaller than about 70 nm \cite{Liu2014}. Under these conditions the transition in a Fe$_3$O$_4$ film is no longer sharp, with the broadness determined by the distribution of the crystallite or domain sizes in the film. For 5 nm Fe$_3$O$_4$ films and thinner Liu \textit{et al.} did not observe a Verwey transition at all \cite{Liu2014}. For our heterostructure, the disordered RHEED patterns, with the presence of broad spots, can be an indication of reduced structural domain sizes. It is therefore conceivable that the subpar quality of the magnetite layer does not allow for the Verwey transition to occur. 

Figure \ref{Fig5}(b) shows the comparative magnetoconductance measurements for a reference 10 QL Bi$_2$Te$_3$ film grown on Al$_2$O$_3$ (0001) (top) and the heterostructure of 40 nm Fe$_3$O$_4$ / 10 QL Bi$_2$Te$_3$ / Al$_2$O$_3$ (0001) (bottom). The weak anti-localization (WAL), characteristic of TIs, is expected to dominate the magnetoconductance for low temperatures and low magnetic fields. Additionally, for the heterostructures containing a magnetic layer, the weak localization (WL) is expected to arise as a signature of a gap opening by magnetic ordering \cite{Lu2011, Yang2019}. This behavior can be described by an approximation of the Hikami-Larkin-Nagaoka (HLN) formula \cite{Hikami1980}, given by

\begin{equation}
	\Delta G_{xx}=\alpha \frac{e^2}{\pi h}\left[\ln\left(\frac{B_{\phi}}{B}\right)-\psi \left(\frac{1}{2}+\frac{B_{\phi}}{B}\right)\right]+\beta B^2
	\label{HLN_eq}
\end{equation}

where $\Delta G_{xx} = G_{xx}(B) - G_{xx}(0)$, $\alpha \equiv \alpha_0 + \alpha_1$ is a pre-factor which describes both WL ($\alpha_0 < 0$) and WAL ($\alpha_1 = 1/2$ per independent topological transport channel), $B_{\phi} = h/(8 \pi e l_{\phi}^2)$, $B$ is the applied magnetic field, $l_{\phi}$ is the phase coherence length, $\psi$ is the digamma function and $\beta$ is the coefficient of the magnetic field. $\alpha$, $l_{\phi}$ and $\beta$ are used as fitting parameters of the HLN equation. In this report, we use one set of $\alpha$ and $l_{\phi}$ for all the fits.

The magnetoconductance for the reference sample grown on Al$_2$O$_3$ (0001), Figure \ref{Fig5}(b)-top, shows the typical behavior for a TI. The pronounced cusp at low temperatures and low magnetic fields shows the predominance of the WAL effect and it can be fitted by equation (\ref{HLN_eq}).
The results of the fit at different temperatures for the reference sample are shown in Figure \ref{Fig5}(c). $\alpha = \alpha_1$ has the expected value of $\approx 0.5$ for the lowest measured temperature, indicating the presence of one conducting channel, since the top and bottom conducting channels are coupled through the bulk carriers in the thin layer \cite{Brahlek2014}. The decreasing value of $\alpha$ with temperature has been previously reported \cite{Kandala2013, Jiang2014} and is usually attributed to thermal broadening. The phase coherence length has a value of $\approx$ 285 nm at 2 K, which is similar to previous studies on TI thin films \cite{He2011, Kandala2013, Jiang2014}. The dependence of $l_{\phi}$ with temperature is shown in the inset in Figure \ref{Fig5}(c). Theoretically, the coherence length is proportional to $T^{-1/2}$ for the two-dimensional system and $T^{-1/3}$ for the one-dimensional system if one considers an inelastic electron-electron scattering mechanism \cite{Altshuler1982}. Our fit is therefore very close to the 2D case, with $l_{\phi} \approx T^{-0.46 \pm 0.01}$.

For the magnetoconductance of the heterostructure plotted in Figure \ref{Fig5}(b)-bottom, however, there is no apparent dependence on the temperature and the typical WAL cusp is absent. In fact, the magnetoconductance now displays only a parabolic $B$-field dependence. Previously reported experiments show that the WAL effect was completely quenched for 1 ML Fe deposited on Bi$_2$Te$_3$ thin films \cite{He2011}. However, this behavior cannot be unambiguously attributed to a gap opening due to the proximity effect, since random magnetic scattering can cause a similar effect \cite{Lu2011}. 

The latter is indeed a very plausible scenario for the heterostructure of 40 nm Fe$_3$O$_4$/10 QL Bi$_2$Te$_3$/Al$_2$O$_3$ (0001). Despite a careful optimization of the growth process, we were not able to prevent a substantial intermixing of iron and the TI layer, which has also been reported in an earlier study \cite{Sanchez-Barriga2017}. Such a scenario could explain a $B^2$-dependence of the magnetoconductance, as seen in Figure \ref{Fig5}(b)-bottom.

\subsection{Bi$_2$Te$_3$ on Fe$_3$O$_4$ on MgO (001)}

\begin{figure}[htbp]
\includegraphics[width=\columnwidth]{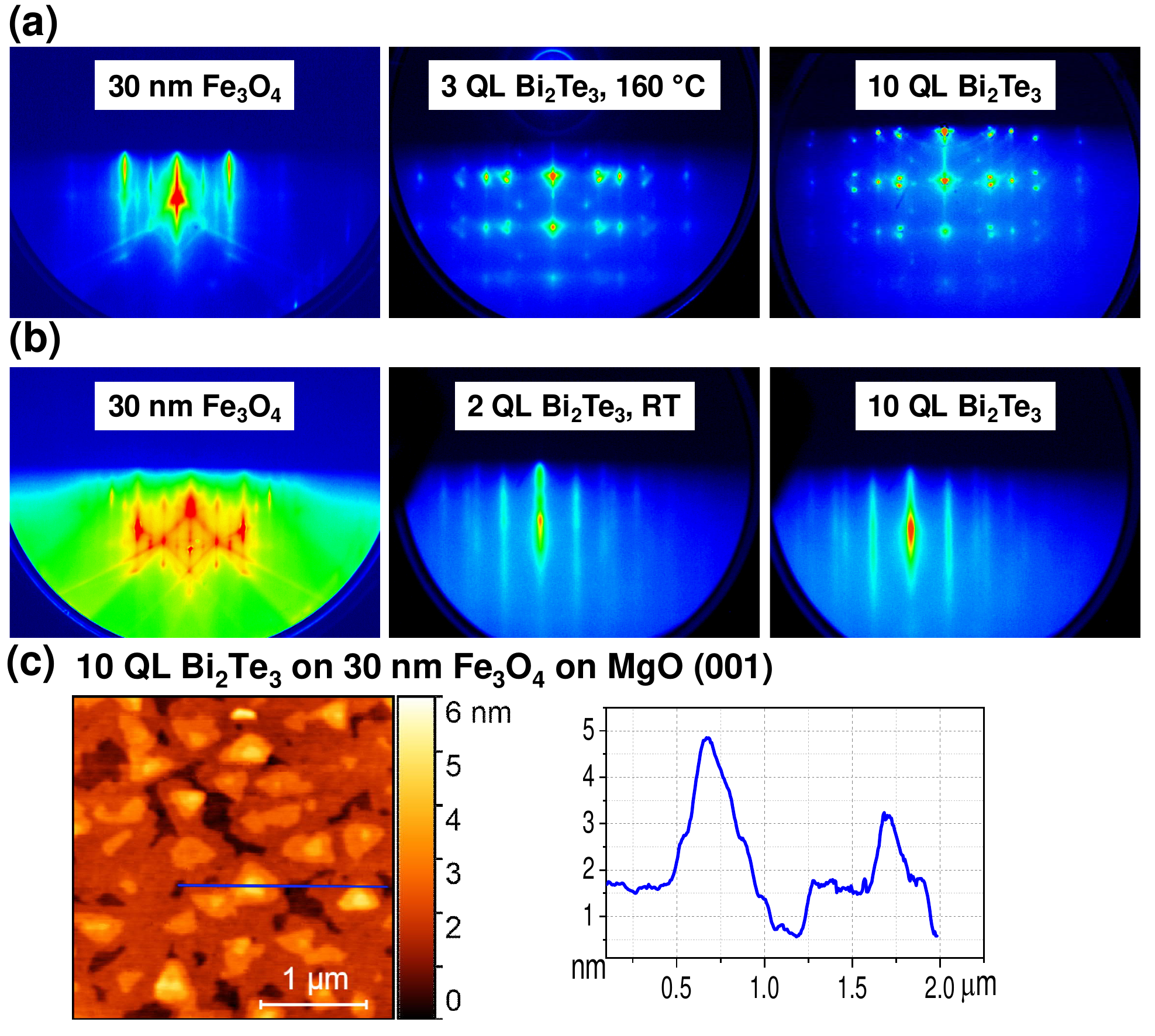}
\caption{RHEED patterns of the heterostructures of Bi$_2$Te$_3$ grown on 30 nm of Fe$_3$O$_4$ films where (a) the first 3 QLs of the topological insulator layer were grown at 160 $^\circ$C; (b) the first 2 QLs were grown at room temperature. In both cases, the first layers were annealed at 240 $^\circ$C in Te atmosphere and the following layers were grown at 220 $^\circ$C. (c) Morphological characterization by AFM of 10 QLs of Bi$_2$Te$_3$ on 30 nm of Fe$_3$O$_4$ on MgO (001) grown under the conditions described in (b).} 
\label{Fig6}
\end{figure}

The absence of the WAL effect in the samples described in the previous section motivated a different approach when interfacing magnetite and Bi$_2$Te$_3$. The current section reports on heterostructures of Bi$_2$Te$_3$ grown on Fe$_3$O$_4$ (001) on MgO (001). For all the films, the magnetite layer has a thickness of 30 nm, which shows a sharp RHEED pattern with the presence of Kikuchi lines and the $\left(\sqrt{2}\times\sqrt{2}\right)R45^\circ$ surface reconstruction signature \cite{Liu2014}. Following this, the growth of the Bi$_2$Te$_3$ layer was carried out in a two-step procedure. Attempts to grow the topological insulator layer at a substrate temperature of 160 $^\circ$C in the first step led to poor crystalline quality, as can be seen in Figure \ref{Fig6} (a). The RHEED shows polycrystalline and island growth that does not improve even for the thicker film, with 10 QLs. The two-step procedure was then adapted, in which the first 2 QLs were grown at room temperature and, after the annealing, the subsequent QLs were grown at 220 $^\circ$C.  The result is displayed in Figure \ref{Fig6} (b). It is remarkable that even for a 2 QL film, the RHEED pattern shows streaky lines, indicative of the good crystalline order of the film. These become clearer in the 10 QL film. However, the film also displays additional streaks when compared to the TI grown on Al$_2$O$_3$ (cf. Figure \ref{Fig4}(a)), which is an indication of the presence of multiple domains.

\begin{figure*}[htbp]
\includegraphics[width=\textwidth]{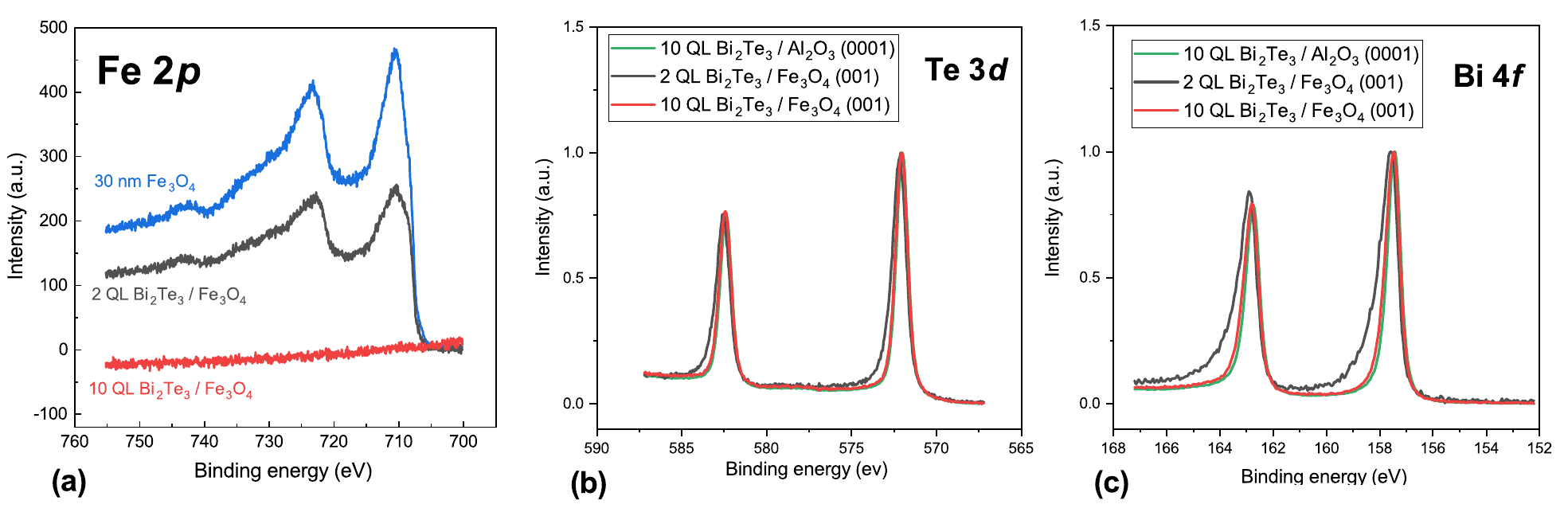}
\caption{XPS spectra of Fe 2\textit{p} (a), Te 3\textit{d} (b) and Bi 4\textit{f} (c) core levels for each step of the growth of 10 QLs Bi$_2$Te$_3$ on 30 nm Fe$_3$O$_4$ on MgO (001). For (b) and (c), the reference spectra of 10 QLs Bi$_2$Te$_3$ on Al$_2$O$_3$ (0001) are shown in green.} 
\label{Fig7}
\end{figure*}

\begin{figure}
\includegraphics[width=\columnwidth]{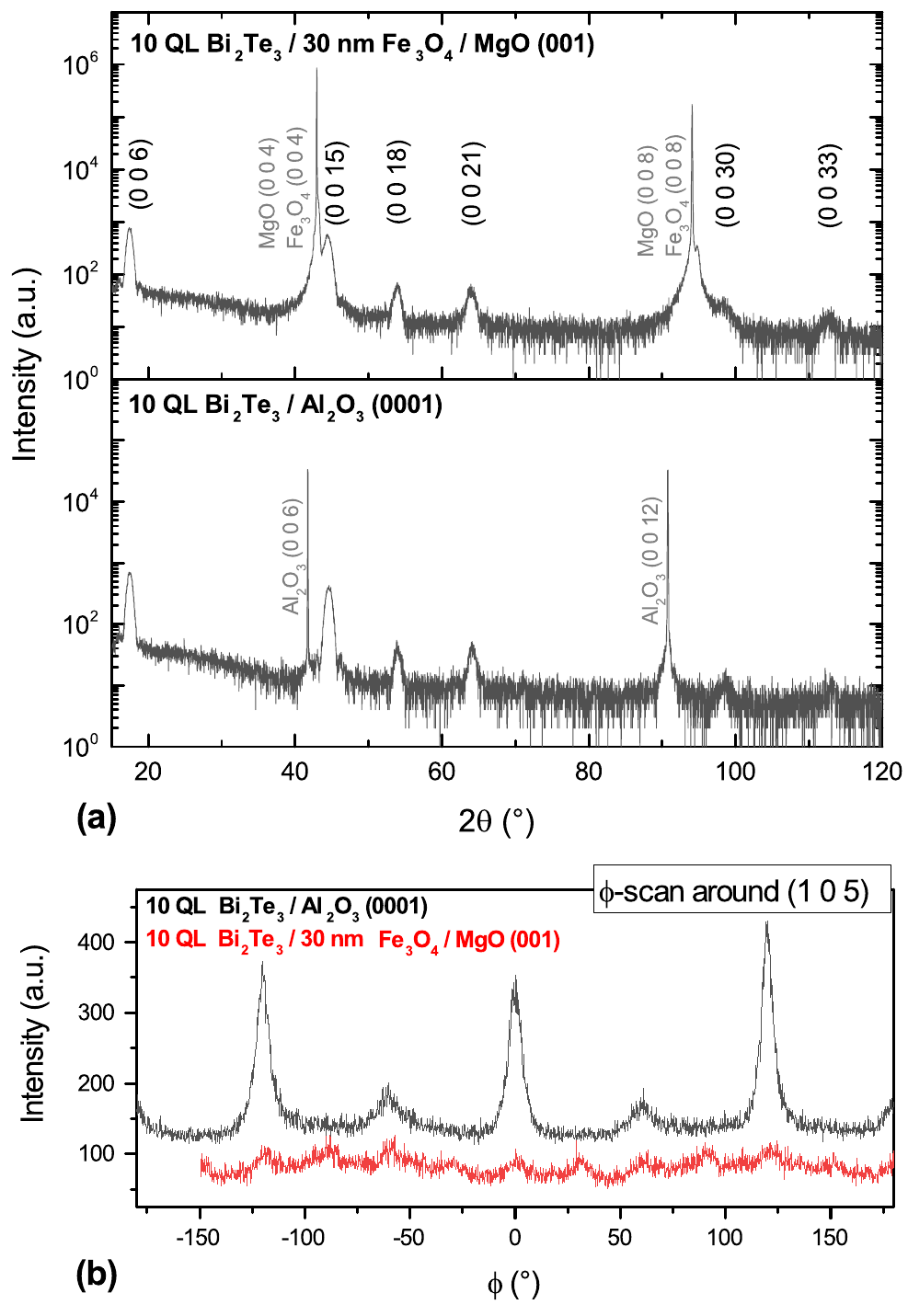}
\caption{(a) XRD $\theta-2\theta$ scans of the heterostructure 10 QL Bi$_2$Te$_3$ / 30 nm Fe$_3$O$_4$ / MgO (001) (\textit{top}) and the reference grown on Al$_2$O$_3$ (00001) (\textit{bottom}). (b) In-plane $\phi$-rotation around the (1 0 5) Bi$_2$Te$_3$ peak for the same samples. An increased in-plane disorder can be noticed for the heterostructure.}
\label{Fig8}
\end{figure}

\begin{figure}
\includegraphics[width=\columnwidth]{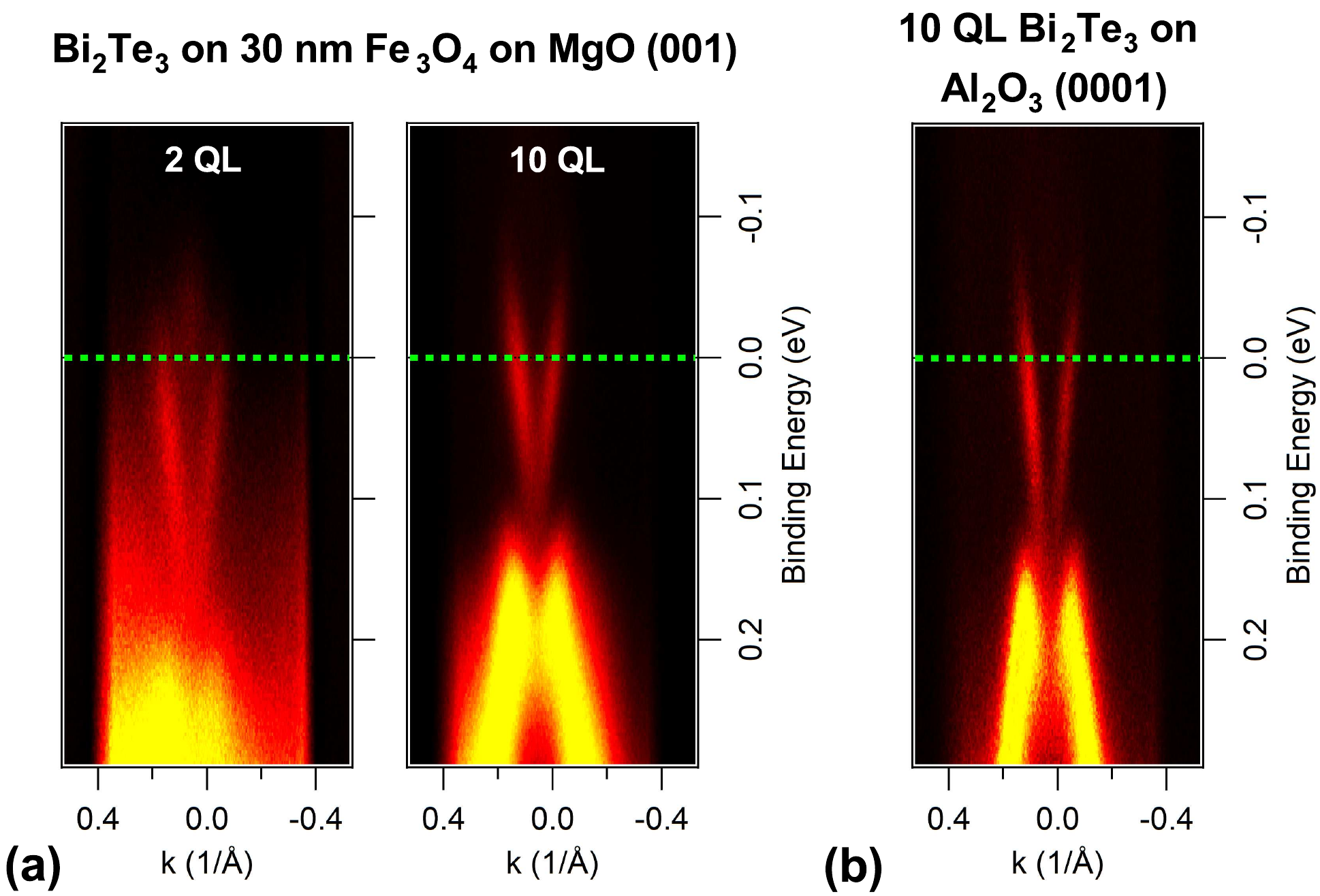}
\caption{\textit{In situ} ARPES spectra of 2 QL and 10 QL Bi$_2$Te$_3$ on 30 nm Fe$_3$O$_4$ on MgO (001) (a) and spectra of a reference sample grown on Al$_2$O$_3$ (0001) (b).}
\label{Fig9}
\end{figure}

\begin{figure*}[htbp]
\includegraphics[width=\textwidth]{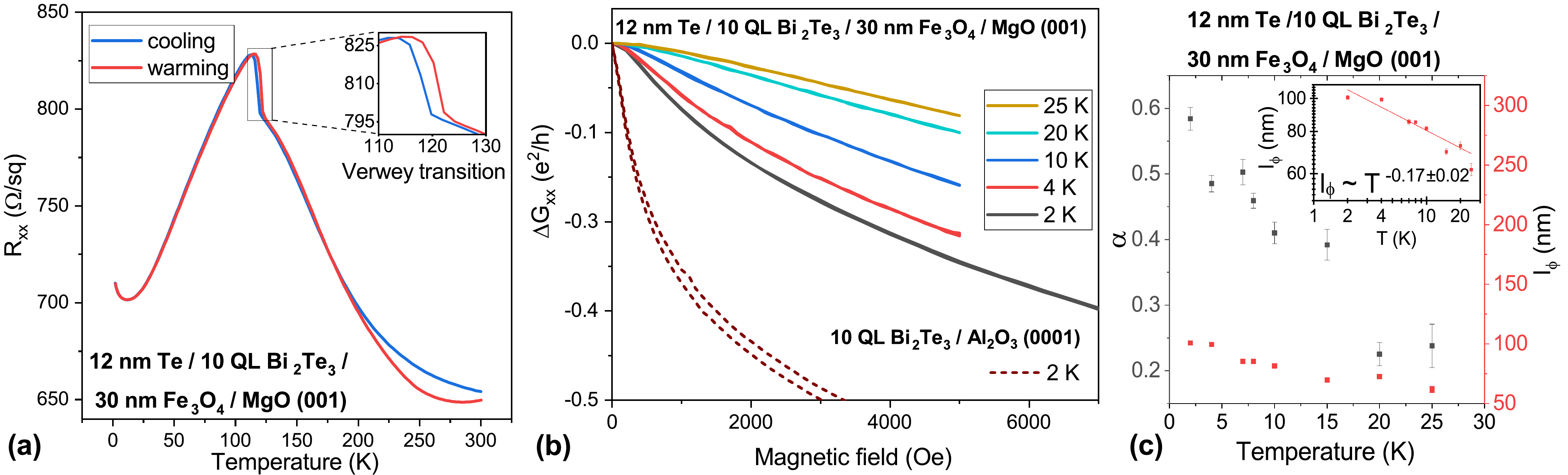}
\caption{(a) Sheet resistance as a function of temperature for the optimized heterostructure of 10 QL Bi$_2$Te$_3$ on top of 30 nm Fe$_3$O$_4$ on MgO (001). (b) Magnetoconductance for the same heterostructure. The data taken at 2 K for the reference sample grown on Al$_2$O$_3$ (0001) are plotted in dashed line for visual comparison. (c) Dependence of the HLN fitting parameters, $\alpha$ and $l_{\phi}$, for the same heterostructure. } 
\label{Fig10}
\end{figure*}

In order to investigate the quality of the interface between Fe$_3$O$_4$ and Bi$_2$Te$_3$, XPS measurements were performed for all the steps of the growth process. Figure \ref{Fig7} shows the Fe 2\textit{p}, Te 3\textit{d} and Bi 4\textit{f} XPS spectra. For the Fe 2\textit{p} peak, one can observe that the signal is quite reduced when we have 2 QLs of Bi$_2$Te$_3$, and disappears for 10 QLs. This implies that a relatively uniform, closed layer of the topological insulator is indeed covering the magnetite. Regarding the Te 3\textit{d} and Bi 4\textit{f} core levels for the 2 QL film, a noticeable shoulder appears at higher binding energies. This indicates a reaction between the layers, compatible with the formation of Bi-O and Te-O bonds. Nevertheless, no strong signal of metallic bismuth or tellurium and bismuth oxides is visible. This is to be contrasted to what was reported in the previous section, where even for the optimized heterostructure, peaks indicative of oxides could be identified. Finally, for the thicker film, the line shape is similar to that of the Bi$_2$Te$_3$ reference grown on Al$_2$O$_3$ (0001). 

The morphological characterization performed by AFM of a 10 QL Bi$_2$Te$_3$ film grown on 30 nm Fe$_3$O$_4$ on MgO (001) is presented in Figure \ref{Fig6}(c). Similarly to the film grown on Al$_2$O$_3$ (0001) (cf. Figure \ref{Fig4}(c)) the pyramidal structure with steps of 1 QL-height are visible. However there are an increased number of rotated domains in-plane. This is in agreement with the RHEED measurements, which indicate an increase in the in-plane disorder of the film. 

From the XRD scans, Figure \ref{Fig8}(a), one can observe that for both the TI grown on Al$_2$O$_3$ (0001) and on 30 nm Fe$_3$O$_4$ / MgO (001), all the peaks can be identified as the (0 0 n) family plane of the Bi$_2$Te$_3$ phase or the underlying layers, implying a good orientation along the c-direction. From the in-plane $\phi$-rotation scan around the (1 0 5) Bi$_2$Te$_3$ peak in Figure \ref{Fig8}(b), however, it is possible to observe significant differences between the heterostructure and the reference sample. From the TI/Al$_2$O$_3$ (0001) scan, one can observe the predominance of one domain with three-fold symmetry and a second domain with significantly less intensity. This is to be expected for Bi$_2$Te$_3$ grown on Al$_2$O$_3$ (0001) due to the large lattice mismatch, which favors the formation of domains with 60$^\circ$ rotation \cite{Hoefer2016}. For the TI grown on the magnetite film, on the other hand, the $\phi$-scan shows a higher degree of in-plane disorder, with weak reflections occurring every 30$^\circ$. The lattice mismatch and the different symmetries between Fe$_3$O$_4$ (001) and Bi$_2$Te$_3$ increase the number of rotated domains around the c-axis. The XRD results are consistent with the RHEED patterns and AFM measurements displayed in Figures \ref{Fig4}(a,c) and \ref{Fig6}(b,c).

Figure \ref{Fig9}(a) shows the ARPES spectra for the two steps of growth of the TI on a magnetite film, as well as a reference spectra for 10 QLs of Bi$_2$Te$_3$ on Al$_2$O$_3$ (b). Even for nominally 2 QLs of Bi$_2$Te$_3$ the surface states are visible, albeit on top of a strong background and a visible contribution from the bulk conduction band. Thickness dependent studies on Bi$_2$Te$_3$ thin films have shown very similar results \cite{Li2010}. The topological features start to appear for a thickness of 2 nm, and the contribution from the bulk conduction band becomes increasingly reduced as the thickness of the film increases. This is also observable in our films and, for 10 QLs, the surface states are now clearly visible and intersect the Fermi level without any contribution from the bulk. The spectrum is similar to the reference (b) and the position of the Dirac point ($\approx$ 150 meV) does not present significant changes, indicating a conservation of the top topological surface states. Nevertheless, the spectrum regarding the heterostructure appears more blurred, which is, once again, consistent with the presence of rotated domains around the c-axis.

The temperature-dependent sheet resistance depicted in Figure \ref{Fig10}(a) shows now the expected behavior for a parallel resistance between the layers: for high temperatures, the resistance increases as the temperature decreases, as characteristic for Fe$_3$O$_4$. At $T \approx T_{\rm{V}} \approx$ 120 K, the Verwey transition is visible in the form of a jump in the resistivity and, for $T < T_{\rm{V}}$, the transport is dominated by the more conductive Bi$_2$Te$_3$ layer, with the upturn characteristic of topological insulators at around 10 K.    

Figure \ref{Fig10}(b) displays the magnetoconductance for a film with 10 QLs of Bi$_2$Te$_3$ on 30 nm of Fe$_3$O$_4$ on MgO (001). Contrary to the previous heterostructure (Figure \ref{Fig5}(b)), the WAL effect is still present at low temperatures and low magnetic fields. However, a visual comparison with the TI grown on Al$_2$O$_3$ (0001) (dashed line) shows that the WAL feature is suppressed in the case of the magnetic heterostructure. The XPS results in Figure \ref{Fig7} suggest the presence of some chemical reaction at the interface. However, the amount is very small so that we can readily expect that the exchange coupling between Fe$_3$O$_4$ and Bi$_2$Te$_3$ layers will still be intact. Furthermore, it has been reported that WL due to bulk subbands in ultrathin films \cite{Lu2011a} and defect-induced WL \cite{Banerjee2014} can occur. Based on our sample characterization, the quality of the Bi$_2$Te$_3$ layers is comparable for the Al$_2$O$_3$ and Fe$_3$O$_4$ substrates. Therefore, we rule out these effects as the main origin of the suppressed WAL, which thus should likely originate from the magnetic interaction with the Fe$_3$O$_4$ surface.

In order to understand the effect of the proximity with a magnetic layer, the HLN equation (\ref{HLN_eq}) was fitted to the data. The evolution of $\alpha$ and $l_{\phi}$ with the temperature is presented in Figure \ref{Fig10}(c). The strong reduction of the phase coherence length of the heterostructure ($l_{\phi} \approx$ 100 nm at 2 K), when compared to the sample grown on a non-magnetic substrate ($l_{\phi} \approx$ 285 nm at 2 K), can be explained by the additional magnetic scattering due to the proximity to the underlying magnetic layer. Furthermore, the dependence of $l_{\phi}$ with the temperature seems to be altered, being now described by $l_{\phi} \approx T^{-0.17 \pm 0.02}$. The decay of the coherence length of our Bi$_2$Te$_3$/Fe$_3$O$_4$ heterostructure thus deviates significantly from the theoretical model, suggesting that other scattering mechanisms, likely related to the magnetic interactions, play a crucial role. Conversely, the pre-factor $\alpha$ is similar to the reference sample. Reducing the thickness of the TI layer to 6 QLs leads to a suppression of $\alpha$ to a value of 0.39 (Ref. \onlinecite{Pereira2020}). 

These results are compatible with a possible opening of a gap in the surface states at the interface between magnetite and the TI, leading to a competition between the WAL and WL effects \cite{Lu2011, Jiang2014, Yang2019} and ultimately resulting in reduced values of $\alpha$ and $l_{\phi}$.

\section{Conclusions}

From the comparative studies of the growth of heterostructures of Bi$_2$Te$_3$ and Fe$_3$O$_4$, one can conclude that, for the case of Fe$_3$O$_4$ on top of Bi$_2$Te$_3$, we encountered a very narrow growth window. Even for the best conditions, the quality of the film and the interface is less than optimal. The absence of the weak anti-localization effect in the magnetoconductance of this type of heterostructure is likely caused by the chemical and magnetic disorder across the interface.

On the other hand, we were able to obtain good quality films of Bi$_2$Te$_3$ on top of Fe$_3$O$_4$, in which the quality of both layers is comparable to our previous works on the individual materials. The good crystallinity observed by RHEED and the preservation of the top topological surface states observed by ARPES are also promising indications of the high quality of the heterostructures. Furthermore, the magnetoconductance for these heterostructures shows a suppression of the surface transport, resultant from the competition between WAL and WL effects, consistent with a gap opening due to the magnetic proximity effect. 

Our work emphasizes the importance of chemically clean interfaces for the study of ferromagnetism induced by the magnetic proximity effect. The good quality of the Bi$_2$Te$_3$ / Fe$_3$O$_4$ / MgO (001) heterostructure indicates that the magnetic proximity effect can be a viable approach for the introduction of magnetic order in TI systems. The experimental realization of chemically clean interfaces together with the unique characteristics of the heterostructures that allow for a uniform magnetization of the TI pave the way for the experimental observation of the QAHE at higher temperatures than those reported in doped systems.

\begin{acknowledgments}
The authors would like to thank Steffen Wirth for the valuable discussions. We would also like to thank Katharina H{\"o}fer and Christoph Becker for the skillful technical assistance and the department of Claudia Felser for the use of the thin films XRD instrument. Financial support from the DFG through Priority Program SPP-1666 Topological Insulators and the Max Planck-POSTECH-Hsinchu Center for Complex Phase Materials is gratefully acknowledged. C.N.W. acknowledges support from the Ministry of Science and Technology of Taiwan, through grant MoST 105-2112-M-007-014-MY3, and V.M.P. from the International Max Planck Research School for Chemistry and Physics of Quantum Materials (IMPRS-CPQM). 
\\

The data that support the findings of this study are available from the corresponding author upon reasonable request.
\end{acknowledgments}

%\bibliography{Ref}

\begin{thebibliography}{36}%
\makeatletter
\providecommand \@ifxundefined [1]{%
 \@ifx{#1\undefined}
}%
\providecommand \@ifnum [1]{%
 \ifnum #1\expandafter \@firstoftwo
 \else \expandafter \@secondoftwo
 \fi
}%
\providecommand \@ifx [1]{%
 \ifx #1\expandafter \@firstoftwo
 \else \expandafter \@secondoftwo
 \fi
}%
\providecommand \natexlab [1]{#1}%
\providecommand \enquote  [1]{``#1''}%
\providecommand \bibnamefont  [1]{#1}%
\providecommand \bibfnamefont [1]{#1}%
\providecommand \citenamefont [1]{#1}%
\providecommand \href@noop [0]{\@secondoftwo}%
\providecommand \href [0]{\begingroup \@sanitize@url \@href}%
\providecommand \@href[1]{\@@startlink{#1}\@@href}%
\providecommand \@@href[1]{\endgroup#1\@@endlink}%
\providecommand \@sanitize@url [0]{\catcode `\\12\catcode `\$12\catcode
  `\&12\catcode `\#12\catcode `\^12\catcode `\_12\catcode `\%12\relax}%
\providecommand \@@startlink[1]{}%
\providecommand \@@endlink[0]{}%
\providecommand \url  [0]{\begingroup\@sanitize@url \@url }%
\providecommand \@url [1]{\endgroup\@href {#1}{\urlprefix }}%
\providecommand \urlprefix  [0]{URL }%
\providecommand \Eprint [0]{\href }%
\providecommand \doibase [0]{http://dx.doi.org/}%
\providecommand \selectlanguage [0]{\@gobble}%
\providecommand \bibinfo  [0]{\@secondoftwo}%
\providecommand \bibfield  [0]{\@secondoftwo}%
\providecommand \translation [1]{[#1]}%
\providecommand \BibitemOpen [0]{}%
\providecommand \bibitemStop [0]{}%
\providecommand \bibitemNoStop [0]{.\EOS\space}%
\providecommand \EOS [0]{\spacefactor3000\relax}%
\providecommand \BibitemShut  [1]{\csname bibitem#1\endcsname}%
\let\auto@bib@innerbib\@empty
%</preamble>
\bibitem [{\citenamefont {Fu}, \citenamefont {Kane},\ and\ \citenamefont
  {Mele}(2007)}]{Fu2007}%
  \BibitemOpen
  \bibfield  {author} {\bibinfo {author} {\bibfnamefont {L.}~\bibnamefont
  {Fu}}, \bibinfo {author} {\bibfnamefont {C.~L.}\ \bibnamefont {Kane}}, \ and\
  \bibinfo {author} {\bibfnamefont {E.~J.}\ \bibnamefont {Mele}},\ }\href
  {\doibase 10.1103/physrevlett.98.106803} {\bibfield  {journal} {\bibinfo
  {journal} {Physical Review Letters}\ }\textbf {\bibinfo {volume} {98}},\
  \bibinfo {pages} {106803} (\bibinfo {year} {2007})}\BibitemShut {NoStop}%
\bibitem [{\citenamefont {Chang}\ \emph {et~al.}(2013)\citenamefont {Chang},
  \citenamefont {Zhang}, \citenamefont {Feng}, \citenamefont {Shen},
  \citenamefont {Zhang}, \citenamefont {Guo}, \citenamefont {Li}, \citenamefont
  {Ou}, \citenamefont {Wei}, \citenamefont {Wang}, \citenamefont {Ji},
  \citenamefont {Feng}, \citenamefont {Ji}, \citenamefont {Chen}, \citenamefont
  {Jia}, \citenamefont {Dai}, \citenamefont {Fang}, \citenamefont {Zhang},
  \citenamefont {He}, \citenamefont {Wang}, \citenamefont {Lu}, \citenamefont
  {Ma},\ and\ \citenamefont {Xue}}]{Chang2013}%
  \BibitemOpen
  \bibfield  {author} {\bibinfo {author} {\bibfnamefont {C.-Z.}\ \bibnamefont
  {Chang}}, \bibinfo {author} {\bibfnamefont {J.}~\bibnamefont {Zhang}},
  \bibinfo {author} {\bibfnamefont {X.}~\bibnamefont {Feng}}, \bibinfo {author}
  {\bibfnamefont {J.}~\bibnamefont {Shen}}, \bibinfo {author} {\bibfnamefont
  {Z.}~\bibnamefont {Zhang}}, \bibinfo {author} {\bibfnamefont
  {M.}~\bibnamefont {Guo}}, \bibinfo {author} {\bibfnamefont {K.}~\bibnamefont
  {Li}}, \bibinfo {author} {\bibfnamefont {Y.}~\bibnamefont {Ou}}, \bibinfo
  {author} {\bibfnamefont {P.}~\bibnamefont {Wei}}, \bibinfo {author}
  {\bibfnamefont {L.-L.}\ \bibnamefont {Wang}}, \bibinfo {author}
  {\bibfnamefont {Z.-Q.}\ \bibnamefont {Ji}}, \bibinfo {author} {\bibfnamefont
  {Y.}~\bibnamefont {Feng}}, \bibinfo {author} {\bibfnamefont {S.}~\bibnamefont
  {Ji}}, \bibinfo {author} {\bibfnamefont {X.}~\bibnamefont {Chen}}, \bibinfo
  {author} {\bibfnamefont {J.}~\bibnamefont {Jia}}, \bibinfo {author}
  {\bibfnamefont {X.}~\bibnamefont {Dai}}, \bibinfo {author} {\bibfnamefont
  {Z.}~\bibnamefont {Fang}}, \bibinfo {author} {\bibfnamefont {S.-C.}\
  \bibnamefont {Zhang}}, \bibinfo {author} {\bibfnamefont {K.}~\bibnamefont
  {He}}, \bibinfo {author} {\bibfnamefont {Y.}~\bibnamefont {Wang}}, \bibinfo
  {author} {\bibfnamefont {L.}~\bibnamefont {Lu}}, \bibinfo {author}
  {\bibfnamefont {X.-C.}\ \bibnamefont {Ma}}, \ and\ \bibinfo {author}
  {\bibfnamefont {Q.-K.}\ \bibnamefont {Xue}},\ }\href {\doibase
  10.1126/science.1234414} {\bibfield  {journal} {\bibinfo  {journal}
  {Science}\ }\textbf {\bibinfo {volume} {340}},\ \bibinfo {pages} {167}
  (\bibinfo {year} {2013})}\BibitemShut {NoStop}%
\bibitem [{\citenamefont {Chang}\ \emph {et~al.}(2015)\citenamefont {Chang},
  \citenamefont {Zhao}, \citenamefont {Kim}, \citenamefont {Zhang},
  \citenamefont {Assaf}, \citenamefont {Heiman}, \citenamefont {Zhang},
  \citenamefont {Liu}, \citenamefont {Chan},\ and\ \citenamefont
  {Moodera}}]{Chang2015}%
  \BibitemOpen
  \bibfield  {author} {\bibinfo {author} {\bibfnamefont {C.-Z.}\ \bibnamefont
  {Chang}}, \bibinfo {author} {\bibfnamefont {W.}~\bibnamefont {Zhao}},
  \bibinfo {author} {\bibfnamefont {D.~Y.}\ \bibnamefont {Kim}}, \bibinfo
  {author} {\bibfnamefont {H.}~\bibnamefont {Zhang}}, \bibinfo {author}
  {\bibfnamefont {B.~A.}\ \bibnamefont {Assaf}}, \bibinfo {author}
  {\bibfnamefont {D.}~\bibnamefont {Heiman}}, \bibinfo {author} {\bibfnamefont
  {S.-C.}\ \bibnamefont {Zhang}}, \bibinfo {author} {\bibfnamefont
  {C.}~\bibnamefont {Liu}}, \bibinfo {author} {\bibfnamefont {M.~H.~W.}\
  \bibnamefont {Chan}}, \ and\ \bibinfo {author} {\bibfnamefont {J.~S.}\
  \bibnamefont {Moodera}},\ }\href {\doibase 10.1038/nmat4204} {\bibfield
  {journal} {\bibinfo  {journal} {Nature Materials}\ }\textbf {\bibinfo
  {volume} {14}},\ \bibinfo {pages} {473} (\bibinfo {year} {2015})}\BibitemShut
  {NoStop}%
\bibitem [{\citenamefont {Mogi}\ \emph {et~al.}(2015)\citenamefont {Mogi},
  \citenamefont {Yoshimi}, \citenamefont {Tsukazaki}, \citenamefont {Yasuda},
  \citenamefont {Kozuka}, \citenamefont {Takahashi}, \citenamefont {Kawasaki},\
  and\ \citenamefont {Tokura}}]{Mogi2015}%
  \BibitemOpen
  \bibfield  {author} {\bibinfo {author} {\bibfnamefont {M.}~\bibnamefont
  {Mogi}}, \bibinfo {author} {\bibfnamefont {R.}~\bibnamefont {Yoshimi}},
  \bibinfo {author} {\bibfnamefont {A.}~\bibnamefont {Tsukazaki}}, \bibinfo
  {author} {\bibfnamefont {K.}~\bibnamefont {Yasuda}}, \bibinfo {author}
  {\bibfnamefont {Y.}~\bibnamefont {Kozuka}}, \bibinfo {author} {\bibfnamefont
  {K.~S.}\ \bibnamefont {Takahashi}}, \bibinfo {author} {\bibfnamefont
  {M.}~\bibnamefont {Kawasaki}}, \ and\ \bibinfo {author} {\bibfnamefont
  {Y.}~\bibnamefont {Tokura}},\ }\href {\doibase 10.1063/1.4935075} {\bibfield
  {journal} {\bibinfo  {journal} {Applied Physics Letters}\ }\textbf {\bibinfo
  {volume} {107}},\ \bibinfo {pages} {182401} (\bibinfo {year}
  {2015})}\BibitemShut {NoStop}%
\bibitem [{\citenamefont {Ye}\ \emph {et~al.}(2015)\citenamefont {Ye},
  \citenamefont {Li}, \citenamefont {Zhu}, \citenamefont {Takeda},
  \citenamefont {Saitoh}, \citenamefont {Wang}, \citenamefont {Pan},
  \citenamefont {Nurmamat}, \citenamefont {Sumida}, \citenamefont {Ji},
  \citenamefont {Liu}, \citenamefont {Yang}, \citenamefont {Liu}, \citenamefont
  {Shen}, \citenamefont {Kimura}, \citenamefont {Qiao},\ and\ \citenamefont
  {Xie}}]{Ye2015}%
  \BibitemOpen
  \bibfield  {author} {\bibinfo {author} {\bibfnamefont {M.}~\bibnamefont
  {Ye}}, \bibinfo {author} {\bibfnamefont {W.}~\bibnamefont {Li}}, \bibinfo
  {author} {\bibfnamefont {S.}~\bibnamefont {Zhu}}, \bibinfo {author}
  {\bibfnamefont {Y.}~\bibnamefont {Takeda}}, \bibinfo {author} {\bibfnamefont
  {Y.}~\bibnamefont {Saitoh}}, \bibinfo {author} {\bibfnamefont
  {J.}~\bibnamefont {Wang}}, \bibinfo {author} {\bibfnamefont {H.}~\bibnamefont
  {Pan}}, \bibinfo {author} {\bibfnamefont {M.}~\bibnamefont {Nurmamat}},
  \bibinfo {author} {\bibfnamefont {K.}~\bibnamefont {Sumida}}, \bibinfo
  {author} {\bibfnamefont {F.}~\bibnamefont {Ji}}, \bibinfo {author}
  {\bibfnamefont {Z.}~\bibnamefont {Liu}}, \bibinfo {author} {\bibfnamefont
  {H.}~\bibnamefont {Yang}}, \bibinfo {author} {\bibfnamefont {Z.}~\bibnamefont
  {Liu}}, \bibinfo {author} {\bibfnamefont {D.}~\bibnamefont {Shen}}, \bibinfo
  {author} {\bibfnamefont {A.}~\bibnamefont {Kimura}}, \bibinfo {author}
  {\bibfnamefont {S.}~\bibnamefont {Qiao}}, \ and\ \bibinfo {author}
  {\bibfnamefont {X.}~\bibnamefont {Xie}},\ }\href {\doibase
  10.1038/ncomms9913} {\bibfield  {journal} {\bibinfo  {journal} {Nature
  Communications}\ }\textbf {\bibinfo {volume} {6}},\ \bibinfo {pages} {8913}
  (\bibinfo {year} {2015})}\BibitemShut {NoStop}%
\bibitem [{\citenamefont {Wei}\ \emph {et~al.}(2013)\citenamefont {Wei},
  \citenamefont {Katmis}, \citenamefont {Assaf}, \citenamefont {Steinberg},
  \citenamefont {Jarillo-Herrero}, \citenamefont {Heiman},\ and\ \citenamefont
  {Moodera}}]{Wei2013}%
  \BibitemOpen
  \bibfield  {author} {\bibinfo {author} {\bibfnamefont {P.}~\bibnamefont
  {Wei}}, \bibinfo {author} {\bibfnamefont {F.}~\bibnamefont {Katmis}},
  \bibinfo {author} {\bibfnamefont {B.~A.}\ \bibnamefont {Assaf}}, \bibinfo
  {author} {\bibfnamefont {H.}~\bibnamefont {Steinberg}}, \bibinfo {author}
  {\bibfnamefont {P.}~\bibnamefont {Jarillo-Herrero}}, \bibinfo {author}
  {\bibfnamefont {D.}~\bibnamefont {Heiman}}, \ and\ \bibinfo {author}
  {\bibfnamefont {J.~S.}\ \bibnamefont {Moodera}},\ }\href {\doibase
  10.1103/physrevlett.110.186807} {\bibfield  {journal} {\bibinfo  {journal}
  {Physical Review Letters}\ }\textbf {\bibinfo {volume} {110}},\ \bibinfo
  {pages} {186807} (\bibinfo {year} {2013})}\BibitemShut {NoStop}%
\bibitem [{\citenamefont {Jiang}\ \emph {et~al.}(2014)\citenamefont {Jiang},
  \citenamefont {Katmis}, \citenamefont {Tang}, \citenamefont {Wei},
  \citenamefont {Moodera},\ and\ \citenamefont {Shi}}]{Jiang2014}%
  \BibitemOpen
  \bibfield  {author} {\bibinfo {author} {\bibfnamefont {Z.}~\bibnamefont
  {Jiang}}, \bibinfo {author} {\bibfnamefont {F.}~\bibnamefont {Katmis}},
  \bibinfo {author} {\bibfnamefont {C.}~\bibnamefont {Tang}}, \bibinfo {author}
  {\bibfnamefont {P.}~\bibnamefont {Wei}}, \bibinfo {author} {\bibfnamefont
  {J.~S.}\ \bibnamefont {Moodera}}, \ and\ \bibinfo {author} {\bibfnamefont
  {J.}~\bibnamefont {Shi}},\ }\href {\doibase 10.1063/1.4881975} {\bibfield
  {journal} {\bibinfo  {journal} {Applied Physics Letters}\ }\textbf {\bibinfo
  {volume} {104}},\ \bibinfo {pages} {222409} (\bibinfo {year}
  {2014})}\BibitemShut {NoStop}%
\bibitem [{\citenamefont {Tang}\ \emph {et~al.}(2017)\citenamefont {Tang},
  \citenamefont {Chang}, \citenamefont {Zhao}, \citenamefont {Liu},
  \citenamefont {Jiang}, \citenamefont {Liu}, \citenamefont {McCartney},
  \citenamefont {Smith}, \citenamefont {Chen}, \citenamefont {Moodera},\ and\
  \citenamefont {Shi}}]{Tang2017}%
  \BibitemOpen
  \bibfield  {author} {\bibinfo {author} {\bibfnamefont {C.}~\bibnamefont
  {Tang}}, \bibinfo {author} {\bibfnamefont {C.-Z.}\ \bibnamefont {Chang}},
  \bibinfo {author} {\bibfnamefont {G.}~\bibnamefont {Zhao}}, \bibinfo {author}
  {\bibfnamefont {Y.}~\bibnamefont {Liu}}, \bibinfo {author} {\bibfnamefont
  {Z.}~\bibnamefont {Jiang}}, \bibinfo {author} {\bibfnamefont {C.-X.}\
  \bibnamefont {Liu}}, \bibinfo {author} {\bibfnamefont {M.~R.}\ \bibnamefont
  {McCartney}}, \bibinfo {author} {\bibfnamefont {D.~J.}\ \bibnamefont
  {Smith}}, \bibinfo {author} {\bibfnamefont {T.}~\bibnamefont {Chen}},
  \bibinfo {author} {\bibfnamefont {J.~S.}\ \bibnamefont {Moodera}}, \ and\
  \bibinfo {author} {\bibfnamefont {J.}~\bibnamefont {Shi}},\ }\href {\doibase
  10.1126/sciadv.1700307} {\bibfield  {journal} {\bibinfo  {journal} {Science
  Advances}\ }\textbf {\bibinfo {volume} {3}},\ \bibinfo {pages} {e1700307}
  (\bibinfo {year} {2017})}\BibitemShut {NoStop}%
\bibitem [{\citenamefont {Yang}\ \emph {et~al.}(2019)\citenamefont {Yang},
  \citenamefont {Fanchiang}, \citenamefont {Chen}, \citenamefont {Tseng},
  \citenamefont {Liu}, \citenamefont {Guo}, \citenamefont {Hong}, \citenamefont
  {Lee},\ and\ \citenamefont {Kwo}}]{Yang2019}%
  \BibitemOpen
  \bibfield  {author} {\bibinfo {author} {\bibfnamefont {S.~R.}\ \bibnamefont
  {Yang}}, \bibinfo {author} {\bibfnamefont {Y.~T.}\ \bibnamefont {Fanchiang}},
  \bibinfo {author} {\bibfnamefont {C.~C.}\ \bibnamefont {Chen}}, \bibinfo
  {author} {\bibfnamefont {C.~C.}\ \bibnamefont {Tseng}}, \bibinfo {author}
  {\bibfnamefont {Y.~C.}\ \bibnamefont {Liu}}, \bibinfo {author} {\bibfnamefont
  {M.~X.}\ \bibnamefont {Guo}}, \bibinfo {author} {\bibfnamefont
  {M.}~\bibnamefont {Hong}}, \bibinfo {author} {\bibfnamefont {S.~F.}\
  \bibnamefont {Lee}}, \ and\ \bibinfo {author} {\bibfnamefont
  {J.}~\bibnamefont {Kwo}},\ }\href {\doibase 10.1103/physrevb.100.045138}
  {\bibfield  {journal} {\bibinfo  {journal} {Physical Review B}\ }\textbf
  {\bibinfo {volume} {100}},\ \bibinfo {pages} {045138} (\bibinfo {year}
  {2019})}\BibitemShut {NoStop}%
\bibitem [{\citenamefont {Wray}\ \emph {et~al.}(2011)\citenamefont {Wray},
  \citenamefont {Xu}, \citenamefont {Xia}, \citenamefont {Hsieh}, \citenamefont
  {Fedorov}, \citenamefont {Hor}, \citenamefont {Cava}, \citenamefont {Bansil},
  \citenamefont {Lin},\ and\ \citenamefont {Hasan}}]{Wray2011}%
  \BibitemOpen
  \bibfield  {author} {\bibinfo {author} {\bibfnamefont {L.~A.}\ \bibnamefont
  {Wray}}, \bibinfo {author} {\bibfnamefont {S.-Y.}\ \bibnamefont {Xu}},
  \bibinfo {author} {\bibfnamefont {Y.}~\bibnamefont {Xia}}, \bibinfo {author}
  {\bibfnamefont {D.}~\bibnamefont {Hsieh}}, \bibinfo {author} {\bibfnamefont
  {A.~V.}\ \bibnamefont {Fedorov}}, \bibinfo {author} {\bibfnamefont {Y.~S.}\
  \bibnamefont {Hor}}, \bibinfo {author} {\bibfnamefont {R.~J.}\ \bibnamefont
  {Cava}}, \bibinfo {author} {\bibfnamefont {A.}~\bibnamefont {Bansil}},
  \bibinfo {author} {\bibfnamefont {H.}~\bibnamefont {Lin}}, \ and\ \bibinfo
  {author} {\bibfnamefont {M.~Z.}\ \bibnamefont {Hasan}},\ }\href {\doibase
  10.1038/nphys1838} {\bibfield  {journal} {\bibinfo  {journal} {Nature
  Physics}\ }\textbf {\bibinfo {volume} {7}},\ \bibinfo {pages} {32} (\bibinfo
  {year} {2011})}\BibitemShut {NoStop}%
\bibitem [{\citenamefont {He}\ \emph {et~al.}(2011)\citenamefont {He},
  \citenamefont {Wang}, \citenamefont {Zhang}, \citenamefont {Sou},
  \citenamefont {Wong}, \citenamefont {Wang}, \citenamefont {Lu}, \citenamefont
  {Shen},\ and\ \citenamefont {Zhang}}]{He2011}%
  \BibitemOpen
  \bibfield  {author} {\bibinfo {author} {\bibfnamefont {H.-T.}\ \bibnamefont
  {He}}, \bibinfo {author} {\bibfnamefont {G.}~\bibnamefont {Wang}}, \bibinfo
  {author} {\bibfnamefont {T.}~\bibnamefont {Zhang}}, \bibinfo {author}
  {\bibfnamefont {I.-K.}\ \bibnamefont {Sou}}, \bibinfo {author} {\bibfnamefont
  {G.~K.~L.}\ \bibnamefont {Wong}}, \bibinfo {author} {\bibfnamefont {J.-N.}\
  \bibnamefont {Wang}}, \bibinfo {author} {\bibfnamefont {H.-Z.}\ \bibnamefont
  {Lu}}, \bibinfo {author} {\bibfnamefont {S.-Q.}\ \bibnamefont {Shen}}, \ and\
  \bibinfo {author} {\bibfnamefont {F.-C.}\ \bibnamefont {Zhang}},\ }\href
  {\doibase 10.1103/physrevlett.106.166805} {\bibfield  {journal} {\bibinfo
  {journal} {Physical Review Letters}\ }\textbf {\bibinfo {volume} {106}},\
  \bibinfo {pages} {166805} (\bibinfo {year} {2011})}\BibitemShut {NoStop}%
\bibitem [{\citenamefont {Honolka}\ \emph {et~al.}(2012)\citenamefont
  {Honolka}, \citenamefont {Khajetoorians}, \citenamefont {Sessi},
  \citenamefont {Wehling}, \citenamefont {Stepanow}, \citenamefont {Mi},
  \citenamefont {Iversen}, \citenamefont {Schlenk}, \citenamefont {Wiebe},
  \citenamefont {Brookes}, \citenamefont {Lichtenstein}, \citenamefont
  {Hofmann}, \citenamefont {Kern},\ and\ \citenamefont
  {Wiesendanger}}]{Honolka2012}%
  \BibitemOpen
  \bibfield  {author} {\bibinfo {author} {\bibfnamefont {J.}~\bibnamefont
  {Honolka}}, \bibinfo {author} {\bibfnamefont {A.~A.}\ \bibnamefont
  {Khajetoorians}}, \bibinfo {author} {\bibfnamefont {V.}~\bibnamefont
  {Sessi}}, \bibinfo {author} {\bibfnamefont {T.~O.}\ \bibnamefont {Wehling}},
  \bibinfo {author} {\bibfnamefont {S.}~\bibnamefont {Stepanow}}, \bibinfo
  {author} {\bibfnamefont {J.-L.}\ \bibnamefont {Mi}}, \bibinfo {author}
  {\bibfnamefont {B.~B.}\ \bibnamefont {Iversen}}, \bibinfo {author}
  {\bibfnamefont {T.}~\bibnamefont {Schlenk}}, \bibinfo {author} {\bibfnamefont
  {J.}~\bibnamefont {Wiebe}}, \bibinfo {author} {\bibfnamefont {N.~B.}\
  \bibnamefont {Brookes}}, \bibinfo {author} {\bibfnamefont {A.~I.}\
  \bibnamefont {Lichtenstein}}, \bibinfo {author} {\bibfnamefont {{\relax
  Ph}.}~\bibnamefont {Hofmann}}, \bibinfo {author} {\bibfnamefont
  {K.}~\bibnamefont {Kern}}, \ and\ \bibinfo {author} {\bibfnamefont
  {R.}~\bibnamefont {Wiesendanger}},\ }\href {\doibase
  10.1103/physrevlett.108.256811} {\bibfield  {journal} {\bibinfo  {journal}
  {Physical Review Letters}\ }\textbf {\bibinfo {volume} {108}},\ \bibinfo
  {pages} {256811} (\bibinfo {year} {2012})}\BibitemShut {NoStop}%
\bibitem [{\citenamefont {Polyakov}\ \emph {et~al.}(2015)\citenamefont
  {Polyakov}, \citenamefont {Meyerheim}, \citenamefont {Crozier}, \citenamefont
  {Gordon}, \citenamefont {Mohseni}, \citenamefont {Roy}, \citenamefont
  {Ernst}, \citenamefont {Vergniory}, \citenamefont {Zubizarreta},
  \citenamefont {Otrokov}, \citenamefont {Chulkov},\ and\ \citenamefont
  {Kirschner}}]{Polyakov2015}%
  \BibitemOpen
  \bibfield  {author} {\bibinfo {author} {\bibfnamefont {A.}~\bibnamefont
  {Polyakov}}, \bibinfo {author} {\bibfnamefont {H.~L.}\ \bibnamefont
  {Meyerheim}}, \bibinfo {author} {\bibfnamefont {E.~D.}\ \bibnamefont
  {Crozier}}, \bibinfo {author} {\bibfnamefont {R.~A.}\ \bibnamefont {Gordon}},
  \bibinfo {author} {\bibfnamefont {K.}~\bibnamefont {Mohseni}}, \bibinfo
  {author} {\bibfnamefont {S.}~\bibnamefont {Roy}}, \bibinfo {author}
  {\bibfnamefont {A.}~\bibnamefont {Ernst}}, \bibinfo {author} {\bibfnamefont
  {M.~G.}\ \bibnamefont {Vergniory}}, \bibinfo {author} {\bibfnamefont
  {X.}~\bibnamefont {Zubizarreta}}, \bibinfo {author} {\bibfnamefont {M.~M.}\
  \bibnamefont {Otrokov}}, \bibinfo {author} {\bibfnamefont {E.~V.}\
  \bibnamefont {Chulkov}}, \ and\ \bibinfo {author} {\bibfnamefont
  {J.}~\bibnamefont {Kirschner}},\ }\href {\doibase 10.1103/physrevb.92.045423}
  {\bibfield  {journal} {\bibinfo  {journal} {Physical Review B}\ }\textbf
  {\bibinfo {volume} {92}},\ \bibinfo {pages} {045423} (\bibinfo {year}
  {2015})}\BibitemShut {NoStop}%
\bibitem [{\citenamefont {S{\'{a}}nchez-Barriga}\ \emph
  {et~al.}(2017)\citenamefont {S{\'{a}}nchez-Barriga}, \citenamefont
  {Ogorodnikov}, \citenamefont {Kuznetsov}, \citenamefont {Volykhov},
  \citenamefont {Matsui}, \citenamefont {Callaert}, \citenamefont {Hadermann},
  \citenamefont {Verbitskiy}, \citenamefont {Koch}, \citenamefont {Varykhalov},
  \citenamefont {Rader},\ and\ \citenamefont {Yashina}}]{Sanchez-Barriga2017}%
  \BibitemOpen
  \bibfield  {author} {\bibinfo {author} {\bibfnamefont {J.}~\bibnamefont
  {S{\'{a}}nchez-Barriga}}, \bibinfo {author} {\bibfnamefont {I.~I.}\
  \bibnamefont {Ogorodnikov}}, \bibinfo {author} {\bibfnamefont {M.~V.}\
  \bibnamefont {Kuznetsov}}, \bibinfo {author} {\bibfnamefont {A.~A.}\
  \bibnamefont {Volykhov}}, \bibinfo {author} {\bibfnamefont {F.}~\bibnamefont
  {Matsui}}, \bibinfo {author} {\bibfnamefont {C.}~\bibnamefont {Callaert}},
  \bibinfo {author} {\bibfnamefont {J.}~\bibnamefont {Hadermann}}, \bibinfo
  {author} {\bibfnamefont {N.~I.}\ \bibnamefont {Verbitskiy}}, \bibinfo
  {author} {\bibfnamefont {R.~J.}\ \bibnamefont {Koch}}, \bibinfo {author}
  {\bibfnamefont {A.}~\bibnamefont {Varykhalov}}, \bibinfo {author}
  {\bibfnamefont {O.}~\bibnamefont {Rader}}, \ and\ \bibinfo {author}
  {\bibfnamefont {L.~V.}\ \bibnamefont {Yashina}},\ }\href {\doibase
  10.1039/c7cp04875k} {\bibfield  {journal} {\bibinfo  {journal} {Physical
  Chemistry Chemical Physics}\ }\textbf {\bibinfo {volume} {19}},\ \bibinfo
  {pages} {30520} (\bibinfo {year} {2017})}\BibitemShut {NoStop}%
\bibitem [{\citenamefont {Verwey}(1939)}]{Verwey1939}%
  \BibitemOpen
  \bibfield  {author} {\bibinfo {author} {\bibfnamefont {E.~J.~W.}\
  \bibnamefont {Verwey}},\ }\href {\doibase 10.1038/144327b0} {\bibfield
  {journal} {\bibinfo  {journal} {Nature}\ }\textbf {\bibinfo {volume} {144}},\
  \bibinfo {pages} {327} (\bibinfo {year} {1939})}\BibitemShut {NoStop}%
\bibitem [{\citenamefont {Liu}\ \emph {et~al.}(2014)\citenamefont {Liu},
  \citenamefont {Rata}, \citenamefont {Chang}, \citenamefont {Komarek},\ and\
  \citenamefont {Tjeng}}]{Liu2014}%
  \BibitemOpen
  \bibfield  {author} {\bibinfo {author} {\bibfnamefont {X.~H.}\ \bibnamefont
  {Liu}}, \bibinfo {author} {\bibfnamefont {A.~D.}\ \bibnamefont {Rata}},
  \bibinfo {author} {\bibfnamefont {C.~F.}\ \bibnamefont {Chang}}, \bibinfo
  {author} {\bibfnamefont {A.~C.}\ \bibnamefont {Komarek}}, \ and\ \bibinfo
  {author} {\bibfnamefont {L.~H.}\ \bibnamefont {Tjeng}},\ }\href {\doibase
  10.1103/physrevb.90.125142} {\bibfield  {journal} {\bibinfo  {journal}
  {Physical Review B}\ }\textbf {\bibinfo {volume} {90}},\ \bibinfo {pages}
  {125142} (\bibinfo {year} {2014})}\BibitemShut {NoStop}%
\bibitem [{\citenamefont {Liu}\ \emph {et~al.}(2016)\citenamefont {Liu},
  \citenamefont {Chang}, \citenamefont {Rata}, \citenamefont {Komarek},\ and\
  \citenamefont {Tjeng}}]{Liu2016}%
  \BibitemOpen
  \bibfield  {author} {\bibinfo {author} {\bibfnamefont {X.~H.}\ \bibnamefont
  {Liu}}, \bibinfo {author} {\bibfnamefont {C.~F.}\ \bibnamefont {Chang}},
  \bibinfo {author} {\bibfnamefont {A.~D.}\ \bibnamefont {Rata}}, \bibinfo
  {author} {\bibfnamefont {A.~C.}\ \bibnamefont {Komarek}}, \ and\ \bibinfo
  {author} {\bibfnamefont {L.~H.}\ \bibnamefont {Tjeng}},\ }\href {\doibase
  10.1038/npjquantmats.2016.27} {\bibfield  {journal} {\bibinfo  {journal} {npj
  Quantum Materials}\ }\textbf {\bibinfo {volume} {1}},\ \bibinfo {pages}
  {16027} (\bibinfo {year} {2016})}\BibitemShut {NoStop}%
\bibitem [{\citenamefont {H{\"o}fer}\ \emph {et~al.}(2014)\citenamefont
  {H{\"o}fer}, \citenamefont {Becker}, \citenamefont {Rata}, \citenamefont
  {Swanson}, \citenamefont {Thalmeier},\ and\ \citenamefont
  {Tjeng}}]{Hoefer2014}%
  \BibitemOpen
  \bibfield  {author} {\bibinfo {author} {\bibfnamefont {K.}~\bibnamefont
  {H{\"o}fer}}, \bibinfo {author} {\bibfnamefont {C.}~\bibnamefont {Becker}},
  \bibinfo {author} {\bibfnamefont {D.}~\bibnamefont {Rata}}, \bibinfo {author}
  {\bibfnamefont {J.}~\bibnamefont {Swanson}}, \bibinfo {author} {\bibfnamefont
  {P.}~\bibnamefont {Thalmeier}}, \ and\ \bibinfo {author} {\bibfnamefont
  {L.~H.}\ \bibnamefont {Tjeng}},\ }\href {\doibase 10.1073/pnas.1410591111}
  {\bibfield  {journal} {\bibinfo  {journal} {Proceedings of the National
  Academy of Sciences}\ }\textbf {\bibinfo {volume} {111}},\ \bibinfo {pages}
  {14979} (\bibinfo {year} {2014})}\BibitemShut {NoStop}%
\bibitem [{\citenamefont {Lu}, \citenamefont {Shi},\ and\ \citenamefont
  {Shen}(2011)}]{Lu2011}%
  \BibitemOpen
  \bibfield  {author} {\bibinfo {author} {\bibfnamefont {H.-Z.}\ \bibnamefont
  {Lu}}, \bibinfo {author} {\bibfnamefont {J.}~\bibnamefont {Shi}}, \ and\
  \bibinfo {author} {\bibfnamefont {S.-Q.}\ \bibnamefont {Shen}},\ }\href
  {\doibase 10.1103/physrevlett.107.076801} {\bibfield  {journal} {\bibinfo
  {journal} {Physical Review Letters}\ }\textbf {\bibinfo {volume} {107}},\
  \bibinfo {pages} {076801} (\bibinfo {year} {2011})}\BibitemShut {NoStop}%
\bibitem [{\citenamefont {Kandala}\ \emph {et~al.}(2013)\citenamefont
  {Kandala}, \citenamefont {Richardella}, \citenamefont {Rench}, \citenamefont
  {Zhang}, \citenamefont {Flanagan},\ and\ \citenamefont
  {Samarth}}]{Kandala2013}%
  \BibitemOpen
  \bibfield  {author} {\bibinfo {author} {\bibfnamefont {A.}~\bibnamefont
  {Kandala}}, \bibinfo {author} {\bibfnamefont {A.}~\bibnamefont
  {Richardella}}, \bibinfo {author} {\bibfnamefont {D.~W.}\ \bibnamefont
  {Rench}}, \bibinfo {author} {\bibfnamefont {D.~M.}\ \bibnamefont {Zhang}},
  \bibinfo {author} {\bibfnamefont {T.~C.}\ \bibnamefont {Flanagan}}, \ and\
  \bibinfo {author} {\bibfnamefont {N.}~\bibnamefont {Samarth}},\ }\href
  {\doibase 10.1063/1.4831987} {\bibfield  {journal} {\bibinfo  {journal}
  {Applied Physics Letters}\ }\textbf {\bibinfo {volume} {103}},\ \bibinfo
  {pages} {202409} (\bibinfo {year} {2013})}\BibitemShut {NoStop}%
\bibitem [{\citenamefont {Fu}(2009)}]{Fu2009}%
  \BibitemOpen
  \bibfield  {author} {\bibinfo {author} {\bibfnamefont {L.}~\bibnamefont
  {Fu}},\ }\href {\doibase 10.1103/physrevlett.103.266801} {\bibfield
  {journal} {\bibinfo  {journal} {Physical Review Letters}\ }\textbf {\bibinfo
  {volume} {103}},\ \bibinfo {pages} {266801} (\bibinfo {year}
  {2009})}\BibitemShut {NoStop}%
\bibitem [{\citenamefont {Liu}, \citenamefont {Hsu},\ and\ \citenamefont
  {Liu}(2013)}]{Liu2013}%
  \BibitemOpen
  \bibfield  {author} {\bibinfo {author} {\bibfnamefont {X.}~\bibnamefont
  {Liu}}, \bibinfo {author} {\bibfnamefont {H.-C.}\ \bibnamefont {Hsu}}, \ and\
  \bibinfo {author} {\bibfnamefont {C.-X.}\ \bibnamefont {Liu}},\ }\href
  {\doibase 10.1103/physrevlett.111.086802} {\bibfield  {journal} {\bibinfo
  {journal} {Physical Review Letters}\ }\textbf {\bibinfo {volume} {111}},\
  \bibinfo {pages} {086802} (\bibinfo {year} {2013})}\BibitemShut {NoStop}%
\bibitem [{\citenamefont {Chang}\ \emph {et~al.}(2016)\citenamefont {Chang},
  \citenamefont {Hu}, \citenamefont {Klein}, \citenamefont {Liu}, \citenamefont
  {Sutarto}, \citenamefont {Tanaka}, \citenamefont {Cezar}, \citenamefont
  {Brookes}, \citenamefont {Lin}, \citenamefont {Hsieh}, \citenamefont {Chen},
  \citenamefont {Rata},\ and\ \citenamefont {Tjeng}}]{Chang2016}%
  \BibitemOpen
  \bibfield  {author} {\bibinfo {author} {\bibfnamefont {C.~F.}\ \bibnamefont
  {Chang}}, \bibinfo {author} {\bibfnamefont {Z.}~\bibnamefont {Hu}}, \bibinfo
  {author} {\bibfnamefont {S.}~\bibnamefont {Klein}}, \bibinfo {author}
  {\bibfnamefont {X.~H.}\ \bibnamefont {Liu}}, \bibinfo {author} {\bibfnamefont
  {R.}~\bibnamefont {Sutarto}}, \bibinfo {author} {\bibfnamefont
  {A.}~\bibnamefont {Tanaka}}, \bibinfo {author} {\bibfnamefont {J.~C.}\
  \bibnamefont {Cezar}}, \bibinfo {author} {\bibfnamefont {N.~B.}\ \bibnamefont
  {Brookes}}, \bibinfo {author} {\bibfnamefont {H.-J.}\ \bibnamefont {Lin}},
  \bibinfo {author} {\bibfnamefont {H.~H.}\ \bibnamefont {Hsieh}}, \bibinfo
  {author} {\bibfnamefont {C.~T.}\ \bibnamefont {Chen}}, \bibinfo {author}
  {\bibfnamefont {A.~D.}\ \bibnamefont {Rata}}, \ and\ \bibinfo {author}
  {\bibfnamefont {L.~H.}\ \bibnamefont {Tjeng}},\ }\href {\doibase
  10.1103/physrevx.6.041011} {\bibfield  {journal} {\bibinfo  {journal}
  {Physical Review X}\ }\textbf {\bibinfo {volume} {6}},\ \bibinfo {pages}
  {041011} (\bibinfo {year} {2016})}\BibitemShut {NoStop}%
\bibitem [{\citenamefont {H{\"o}fer}\ \emph {et~al.}(2015)\citenamefont
  {H{\"o}fer}, \citenamefont {Becker}, \citenamefont {Wirth},\ and\
  \citenamefont {Tjeng}}]{Hoefer2015}%
  \BibitemOpen
  \bibfield  {author} {\bibinfo {author} {\bibfnamefont {K.}~\bibnamefont
  {H{\"o}fer}}, \bibinfo {author} {\bibfnamefont {C.}~\bibnamefont {Becker}},
  \bibinfo {author} {\bibfnamefont {S.}~\bibnamefont {Wirth}}, \ and\ \bibinfo
  {author} {\bibfnamefont {L.~H.}\ \bibnamefont {Tjeng}},\ }\href {\doibase
  10.1063/1.4931038} {\bibfield  {journal} {\bibinfo  {journal} {{AIP}
  Advances}\ }\textbf {\bibinfo {volume} {5}},\ \bibinfo {pages} {097139}
  (\bibinfo {year} {2015})}\BibitemShut {NoStop}%
\bibitem [{\citenamefont {Telesca}\ \emph {et~al.}(2012)\citenamefont
  {Telesca}, \citenamefont {Nie}, \citenamefont {Budnick}, \citenamefont
  {Wells},\ and\ \citenamefont {Sinkovic}}]{Telesca2012}%
  \BibitemOpen
  \bibfield  {author} {\bibinfo {author} {\bibfnamefont {D.}~\bibnamefont
  {Telesca}}, \bibinfo {author} {\bibfnamefont {Y.}~\bibnamefont {Nie}},
  \bibinfo {author} {\bibfnamefont {J.~I.}\ \bibnamefont {Budnick}}, \bibinfo
  {author} {\bibfnamefont {B.~O.}\ \bibnamefont {Wells}}, \ and\ \bibinfo
  {author} {\bibfnamefont {B.}~\bibnamefont {Sinkovic}},\ }\href {\doibase
  10.1103/physrevb.85.214517} {\bibfield  {journal} {\bibinfo  {journal}
  {Physical Review B}\ }\textbf {\bibinfo {volume} {85}},\ \bibinfo {pages}
  {214517} (\bibinfo {year} {2012})}\BibitemShut {NoStop}%
\bibitem [{\citenamefont {Gota}\ \emph {et~al.}(1999)\citenamefont {Gota},
  \citenamefont {Guiot}, \citenamefont {Henriot},\ and\ \citenamefont
  {Gautier-Soyer}}]{Gota1999}%
  \BibitemOpen
  \bibfield  {author} {\bibinfo {author} {\bibfnamefont {S.}~\bibnamefont
  {Gota}}, \bibinfo {author} {\bibfnamefont {E.}~\bibnamefont {Guiot}},
  \bibinfo {author} {\bibfnamefont {M.}~\bibnamefont {Henriot}}, \ and\
  \bibinfo {author} {\bibfnamefont {M.}~\bibnamefont {Gautier-Soyer}},\ }\href
  {\doibase 10.1103/physrevb.60.14387} {\bibfield  {journal} {\bibinfo
  {journal} {Physical Review B}\ }\textbf {\bibinfo {volume} {60}},\ \bibinfo
  {pages} {14387} (\bibinfo {year} {1999})}\BibitemShut {NoStop}%
\bibitem [{\citenamefont {Moulder}\ and\ \citenamefont
  {Chastain}(1992)}]{Moulder1992}%
  \BibitemOpen
  \bibfield  {author} {\bibinfo {author} {\bibfnamefont {J.~F.}\ \bibnamefont
  {Moulder}}\ and\ \bibinfo {author} {\bibfnamefont {J.}~\bibnamefont
  {Chastain}},\ }\href
  {https://www.amazon.com/Handbook-Ray-Photoelectron-Spectroscopy-624755/dp/0962702625?SubscriptionId=AKIAIOBINVZYXZQZ2U3A&tag=chimbori05-20&linkCode=xm2&camp=2025&creative=165953&creativeASIN=0962702625}
  {\emph {\bibinfo {title} {Handbook of {X}-ray {P}hotoelectron {S}pectroscopy:
  {A} {R}eference {B}ook of {S}tandard {S}pectra for {I}dentification and
  {I}nterpretation of {XPS} {D}ata}}}\ (\bibinfo  {publisher} {Physical
  Electronics Division, Perkin-Elmer Corporation},\ \bibinfo {year}
  {1992})\BibitemShut {NoStop}%
\bibitem [{\citenamefont {Eerenstein}\ \emph {et~al.}(2002)\citenamefont
  {Eerenstein}, \citenamefont {Palstra}, \citenamefont {Hibma},\ and\
  \citenamefont {Celotto}}]{Eerenstein2002}%
  \BibitemOpen
  \bibfield  {author} {\bibinfo {author} {\bibfnamefont {W.}~\bibnamefont
  {Eerenstein}}, \bibinfo {author} {\bibfnamefont {T.~T.~M.}\ \bibnamefont
  {Palstra}}, \bibinfo {author} {\bibfnamefont {T.}~\bibnamefont {Hibma}}, \
  and\ \bibinfo {author} {\bibfnamefont {S.}~\bibnamefont {Celotto}},\ }\href
  {\doibase 10.1103/physrevb.66.201101} {\bibfield  {journal} {\bibinfo
  {journal} {Physical Review B}\ }\textbf {\bibinfo {volume} {66}},\ \bibinfo
  {pages} {201101(R)} (\bibinfo {year} {2002})}\BibitemShut {NoStop}%
\bibitem [{\citenamefont {Hikami}, \citenamefont {Larkin},\ and\ \citenamefont
  {Nagaoka}(1980)}]{Hikami1980}%
  \BibitemOpen
  \bibfield  {author} {\bibinfo {author} {\bibfnamefont {S.}~\bibnamefont
  {Hikami}}, \bibinfo {author} {\bibfnamefont {A.~I.}\ \bibnamefont {Larkin}},
  \ and\ \bibinfo {author} {\bibfnamefont {Y.}~\bibnamefont {Nagaoka}},\ }\href
  {\doibase 10.1143/ptp.63.707} {\bibfield  {journal} {\bibinfo  {journal}
  {Progress of Theoretical Physics}\ }\textbf {\bibinfo {volume} {63}},\
  \bibinfo {pages} {707} (\bibinfo {year} {1980})}\BibitemShut {NoStop}%
\bibitem [{\citenamefont {Brahlek}\ \emph {et~al.}(2014)\citenamefont
  {Brahlek}, \citenamefont {Koirala}, \citenamefont {Salehi}, \citenamefont
  {Bansal},\ and\ \citenamefont {Oh}}]{Brahlek2014}%
  \BibitemOpen
  \bibfield  {author} {\bibinfo {author} {\bibfnamefont {M.}~\bibnamefont
  {Brahlek}}, \bibinfo {author} {\bibfnamefont {N.}~\bibnamefont {Koirala}},
  \bibinfo {author} {\bibfnamefont {M.}~\bibnamefont {Salehi}}, \bibinfo
  {author} {\bibfnamefont {N.}~\bibnamefont {Bansal}}, \ and\ \bibinfo {author}
  {\bibfnamefont {S.}~\bibnamefont {Oh}},\ }\href {\doibase
  10.1103/physrevlett.113.026801} {\bibfield  {journal} {\bibinfo  {journal}
  {Physical Review Letters}\ }\textbf {\bibinfo {volume} {113}},\ \bibinfo
  {pages} {026801} (\bibinfo {year} {2014})}\BibitemShut {NoStop}%
\bibitem [{\citenamefont {Altshuler}, \citenamefont {Aronov},\ and\
  \citenamefont {Khmelnitsky}(1982)}]{Altshuler1982}%
  \BibitemOpen
  \bibfield  {author} {\bibinfo {author} {\bibfnamefont {B.~L.}\ \bibnamefont
  {Altshuler}}, \bibinfo {author} {\bibfnamefont {A.~G.}\ \bibnamefont
  {Aronov}}, \ and\ \bibinfo {author} {\bibfnamefont {D.~E.}\ \bibnamefont
  {Khmelnitsky}},\ }\href {\doibase 10.1088/0022-3719/15/36/018} {\bibfield
  {journal} {\bibinfo  {journal} {Journal of Physics C: Solid State Physics}\
  }\textbf {\bibinfo {volume} {15}},\ \bibinfo {pages} {7367} (\bibinfo {year}
  {1982})}\BibitemShut {NoStop}%
\bibitem [{\citenamefont {H{\"o}fer}(2016)}]{Hoefer2016}%
  \BibitemOpen
  \bibfield  {author} {\bibinfo {author} {\bibfnamefont {K.}~\bibnamefont
  {H{\"o}fer}},\ }\emph {\bibinfo {title} {All \textit{in situ} ultra-high
  vacuum study of {Bi$_2$Te$_3$} topological insulator thin films}},\
  \href@noop {} {Ph.D. thesis},\ \bibinfo  {school} {Technische Universit{\"a}t
  Dresden} (\bibinfo {year} {2016})\BibitemShut {NoStop}%
\bibitem [{\citenamefont {Li}\ \emph {et~al.}(2010)\citenamefont {Li},
  \citenamefont {Wang}, \citenamefont {Zhu}, \citenamefont {Liu}, \citenamefont
  {Ye}, \citenamefont {Chen}, \citenamefont {Wang}, \citenamefont {He},
  \citenamefont {Wang}, \citenamefont {Ma}, \citenamefont {Zhang},
  \citenamefont {Dai}, \citenamefont {Fang}, \citenamefont {Xie}, \citenamefont
  {Liu}, \citenamefont {Qi}, \citenamefont {Jia}, \citenamefont {Zhang},\ and\
  \citenamefont {Xue}}]{Li2010}%
  \BibitemOpen
  \bibfield  {author} {\bibinfo {author} {\bibfnamefont {Y.-Y.}\ \bibnamefont
  {Li}}, \bibinfo {author} {\bibfnamefont {G.}~\bibnamefont {Wang}}, \bibinfo
  {author} {\bibfnamefont {X.-G.}\ \bibnamefont {Zhu}}, \bibinfo {author}
  {\bibfnamefont {M.-H.}\ \bibnamefont {Liu}}, \bibinfo {author} {\bibfnamefont
  {C.}~\bibnamefont {Ye}}, \bibinfo {author} {\bibfnamefont {X.}~\bibnamefont
  {Chen}}, \bibinfo {author} {\bibfnamefont {Y.-Y.}\ \bibnamefont {Wang}},
  \bibinfo {author} {\bibfnamefont {K.}~\bibnamefont {He}}, \bibinfo {author}
  {\bibfnamefont {L.-L.}\ \bibnamefont {Wang}}, \bibinfo {author}
  {\bibfnamefont {X.-C.}\ \bibnamefont {Ma}}, \bibinfo {author} {\bibfnamefont
  {H.-J.}\ \bibnamefont {Zhang}}, \bibinfo {author} {\bibfnamefont
  {X.}~\bibnamefont {Dai}}, \bibinfo {author} {\bibfnamefont {Z.}~\bibnamefont
  {Fang}}, \bibinfo {author} {\bibfnamefont {X.-C.}\ \bibnamefont {Xie}},
  \bibinfo {author} {\bibfnamefont {Y.}~\bibnamefont {Liu}}, \bibinfo {author}
  {\bibfnamefont {X.-L.}\ \bibnamefont {Qi}}, \bibinfo {author} {\bibfnamefont
  {J.-F.}\ \bibnamefont {Jia}}, \bibinfo {author} {\bibfnamefont {S.-C.}\
  \bibnamefont {Zhang}}, \ and\ \bibinfo {author} {\bibfnamefont {Q.-K.}\
  \bibnamefont {Xue}},\ }\href {\doibase 10.1002/adma.201000368} {\bibfield
  {journal} {\bibinfo  {journal} {Advanced Materials}\ }\textbf {\bibinfo
  {volume} {22}},\ \bibinfo {pages} {4002} (\bibinfo {year}
  {2010})}\BibitemShut {NoStop}%
\bibitem [{\citenamefont {Lu}\ and\ \citenamefont {Shen}(2011)}]{Lu2011a}%
  \BibitemOpen
  \bibfield  {author} {\bibinfo {author} {\bibfnamefont {H.-Z.}\ \bibnamefont
  {Lu}}\ and\ \bibinfo {author} {\bibfnamefont {S.-Q.}\ \bibnamefont {Shen}},\
  }\href {\doibase 10.1103/physrevb.84.125138} {\bibfield  {journal} {\bibinfo
  {journal} {Physical Review B}\ }\textbf {\bibinfo {volume} {84}},\ \bibinfo
  {pages} {125138} (\bibinfo {year} {2011})}\BibitemShut {NoStop}%
\bibitem [{\citenamefont {Banerjee}\ \emph {et~al.}(2014)\citenamefont
  {Banerjee}, \citenamefont {Son}, \citenamefont {Deorani}, \citenamefont
  {Ren}, \citenamefont {Wang},\ and\ \citenamefont {Yang}}]{Banerjee2014}%
  \BibitemOpen
  \bibfield  {author} {\bibinfo {author} {\bibfnamefont {K.}~\bibnamefont
  {Banerjee}}, \bibinfo {author} {\bibfnamefont {J.}~\bibnamefont {Son}},
  \bibinfo {author} {\bibfnamefont {P.}~\bibnamefont {Deorani}}, \bibinfo
  {author} {\bibfnamefont {P.}~\bibnamefont {Ren}}, \bibinfo {author}
  {\bibfnamefont {L.}~\bibnamefont {Wang}}, \ and\ \bibinfo {author}
  {\bibfnamefont {H.}~\bibnamefont {Yang}},\ }\href {\doibase
  10.1103/physrevb.90.235427} {\bibfield  {journal} {\bibinfo  {journal}
  {Physical Review B}\ }\textbf {\bibinfo {volume} {90}},\ \bibinfo {pages}
  {235427} (\bibinfo {year} {2014})}\BibitemShut {NoStop}%
\bibitem [{\citenamefont {Pereira}\ \emph {et~al.}(2020)\citenamefont
  {Pereira}, \citenamefont {Altendorf}, \citenamefont {Liu}, \citenamefont
  {Liao}, \citenamefont {Komarek}, \citenamefont {Guo}, \citenamefont {Lin},
  \citenamefont {Chen}, \citenamefont {Hong}, \citenamefont {Kwo},
  \citenamefont {Tjeng},\ and\ \citenamefont {Wu}}]{Pereira2020}%
  \BibitemOpen
  \bibfield  {author} {\bibinfo {author} {\bibfnamefont {V.~M.}\ \bibnamefont
  {Pereira}}, \bibinfo {author} {\bibfnamefont {S.~G.}\ \bibnamefont
  {Altendorf}}, \bibinfo {author} {\bibfnamefont {C.~E.}\ \bibnamefont {Liu}},
  \bibinfo {author} {\bibfnamefont {S.~C.}\ \bibnamefont {Liao}}, \bibinfo
  {author} {\bibfnamefont {A.~C.}\ \bibnamefont {Komarek}}, \bibinfo {author}
  {\bibfnamefont {M.}~\bibnamefont {Guo}}, \bibinfo {author} {\bibfnamefont
  {H.~J.}\ \bibnamefont {Lin}}, \bibinfo {author} {\bibfnamefont {C.~T.}\
  \bibnamefont {Chen}}, \bibinfo {author} {\bibfnamefont {M.}~\bibnamefont
  {Hong}}, \bibinfo {author} {\bibfnamefont {J.}~\bibnamefont {Kwo}}, \bibinfo
  {author} {\bibfnamefont {L.~H.}\ \bibnamefont {Tjeng}}, \ and\ \bibinfo
  {author} {\bibfnamefont {C.~N.}\ \bibnamefont {Wu}},\ }\href@noop {}
  {\bibfield  {journal} {\bibinfo  {journal} {Physical Review Materials,
  accepted}\ } (\bibinfo {year} {2020})}\BibitemShut {NoStop}%
\end{thebibliography}

\end{document}